\documentclass{emulateapj}

\shorttitle{Star clusters in M82}
\shortauthors{Mayya et al.}

\newcommand{\ha}{H$\alpha$ }
\newcommand{\msun}{$M\odot$}

\begin{document}

\title{HST/ACS imaging of M82: A comparison of mass and size distribution 
functions of the younger nuclear and older disk clusters
}

\author{Y.\ D.\ Mayya, R.\ Romano, L.\ H.\ Rodr\'{\i}guez-Merino, A.\ Luna, 
L.\ Carrasco, and D.\ Rosa-Gonz\'alez}

\affil{Instituto Nacional de Astrofisica, Optica y Electronica, Luis Enrique 
Erro 1, Tonantzintla, C.P. 72840, Puebla, Mexico.}
\accepted{in Astrophysical Journal}

\begin{abstract}
We present the results obtained from an objective search for stellar 
clusters, both in the currently active nuclear starburst region, and in the 
post-starburst disk of M82. Images obtained with the Advanced Camera for 
Surveys (ACS) aboard the {\it Hubble Space Telescope} (HST) in F435W(B), 
F555W(V), and F814W(I) filters were used in the  search for the clusters. We 
detected 653 clusters of which 393 are located outside the central 450~pc in 
the post-starburst disk of M82. The luminosity function of the detected 
clusters show an apparent turnover at B=22~mag ($M_{\rm B}=-5.8$), which we
interpret from Monte Carlo simulations as due to incompleteness in the 
detection of faint clusters, rather than an intrinsic log-normal distribution. 
We derived a photometric mass of every detected cluster from models of 
simple stellar populations assuming a mean age of either an 8 (nuclear 
clusters) or 100 (disk clusters) million years old. The mass functions of the 
disk (older) and the nuclear (younger) clusters follow 
power-laws, 
the former being marginally flatter ($\alpha=1.5\pm0.1$) than the 
latter ($\alpha=1.8\pm0.1$). The distribution of sizes (Full Width at Half 
Maximum) of clusters brighter than the apparent turn-over magnitude 
(mass$\gtrsim2\times10^4$\msun) can be described by a log-normal function.
This function peaks at 10~pc for clusters more massive than $10^5$~\msun, 
whereas for lower masses, the peak is marginally shifted to larger values for 
the younger, and smaller values for the older clusters. The observed 
trend towards 
flattening of the mass function with age, together with an 
over-abundance of older 
compact clusters, imply that cluster disruption in M82 is both dependent on 
the mass and size of the clusters.
\end{abstract}

\section{Introduction}
The high spatial resolution of the Hubble Space Telescope (HST) observations 
have allowed the detection of compact star clusters in starburst regions 
of several galaxies \citep{Hol92,Whi93,Oco95,Whi99,deG03c,Lar04}.
The most massive of them, often referred to as Super Star Clusters (SSCs), 
have masses and sizes comparable to those of galactic globular clusters (GCs). 
The SSCs span a wide range of ages: those associated with currently active 
star forming regions are as young as a few million years old, while the 
others are as old as 500~Myr \citep{Whi99}. Ever since their discovery by 
the HST-based observations, they have been thought to be progenitors of GCs. 
There is little doubt about the survival of the  most massive of the SSCs
over a Hubble time, but an interesting question is what 
fraction of them will survive?

Some star clusters are relatively weakly bound objects, and vulnerable to
disruption by a variety of processes that operate on three different 
timescales \citep{Fal01, Mai04, deG07}. 
On short timescales ($t\sim10^7$~yr), when the clusters and
protoclusters are partly gaseous, the exploding supernovae and the resulting
superwinds remove gaseous mass from clusters leading to cluster expansion 
and disruption, a process popularly dubbed as infant mortality.
On intermediate timescales ($10^7<t<{\rm few} \times 10^8$~yr), 
the mass-loss from evolving stars leads to the disruption of the clusters.
On even longer timescales ($t> {\rm few} \times 10^8$~yr),
stellar dynamical processes, especially evaporation due to two-body scattering,
and tidal effects on a cluster as it orbits around the galaxy, known as
gravitational shocks, come into play in the removal of stellar mass from 
clusters. The first two processes are directly related to the evolution of
massive and intermediate-age stars, and hence their timescales are less 
uncertain. On the other hand, the exact 
timescale for the gravitational shocks is difficult to calculate given that
it depends on factors external to the cluster, such as a detailed modeling 
of the local gravitational potential.
Another external process that can play a role in disrupting clusters is 
their encounters with giant molecular clouds (GMCs) \citep{Gie06b}. The 
timescale for 
this process depends on the molecular gas content and is typically less than 
that for the gravitational shocks in gas-rich environments.
The GCs have already gone through these processes, whereas the SSCs 
are at the right age to see the different disruption processes at work.

How does the various cluster disruption processes affect the cluster
mass distribution function? \citet{Fal01} have addressed this issue in
detail and found that as long as there is no dependence of the fractional
gas mass or the stellar initial mass function (IMF) on the cluster mass,
the first two disruption processes discussed above are not expected to 
affect the cluster mass function (CMF). On the other hand, disruption due 
to gravitational shocks and encounters of clusters with GMCs
is capable of changing the cluster mass function. 
Luminosity and mass functions of GCs follow log-normal distributions peaking 
at $M_V=-7.3$~mag and $2\times10^5$~\msun, respectively \citep{Har91}. 
Meanwhile, young SSCs are found to obey a power-law distribution of 
luminosities ($dN/dL = L^{-\alpha}$ with $\alpha=2.0$) over a wide range 
of cluster luminosities \citep{deG03c}. \citet{Fal01} found that a 
log-normal or quasi log-normal distribution function can be
obtained from an initial power-law type of distribution, through selective 
removal of low-mass clusters from the initial distribution as the cluster 
population evolves. They propose that the SSCs can still be progenitors 
of GCs in spite of them having different functional forms of mass distribution.
The clue to understand the evolution of SSCs towards GCs lies in establishing
the  CMFs of intermediate age clusters. Such studies have been carried out 
for the clusters in the Antennae galaxy \citep{Whi99}, M51 \citep{Gie06a}, 
and region B of M82 \citep{deG03a}. These studies are consistent
with a power-law CMF that is truncated at cluster masses of $\sim10^5$~\msun,
giving the appearance of a quasi-log-normal distribution as predicted by
the \citet{Fal01} scheme. However, difficulties in determining reliable 
ages of slightly evolved clusters, combined with the incompleteness 
affecting the mass function for low cluster masses, have hindered the 
interpretation of the observed mass functions. 

Recent observations of M82 with the HST/ACS instrument, covering the entire
optical extent of the galaxy, offers an excellent opportunity to 
investigate the role of disruption processes on mass function. Being the 
nearest galaxy with a large population of SSCs, the luminosity and size 
distribution functions can be studied better in this galaxy than in any 
other galaxy. In addition, younger and older populations are spatially 
segregated: present starburst activity (age$\lesssim10$~Myr) is rather 
exclusive to the central zone \citep{Rie93,For03}, whereas the disk lacks 
any recent star formation. Almost all the star formation
in the disk took place in a violent disk-wide burst about 100-500~Myr ago,
following the interaction of M82 with the members of M81 group
\citep{May06}. Cluster formation is known to be efficient during the
burst phase of star formation \citep{Bas05}, and hence we expect large 
number of clusters of age $\sim100$~Myr in the disk.
Recent determinations of age of the clusters in M82B region 
by \citet{Smi07} confirm that the peak epoch of cluster formation 
occurred $\sim150$~Myr ago.
The presence of two distinct epochs of cluster formation, 
well separated spatially from each other, makes M82 an ideal candidate
for
a study of evolutionary effects on the cluster mass function.

In \S~2, we give a brief summary of the observational material used in this
study, strategies used for cluster selection, and the analysis carried out
for obtaining their size and luminosities. There, the observed luminosity 
and size distribution functions are also presented.
Results of the cluster simulation are presented in \S~3.
Methods for deriving the mass of individual clusters, and the 
construction of the mass distribution function, 
is described in \S~4. In \S~5, we compare the observed mass and size 
distribution functions in M82 with the functions obtained for other galaxies,
and discuss the most important disruption process active during the first
few $\times10^8$ yr in M82. The conclusions from this study are summarized 
in \S~6.

\section{Observations, source selection and analysis}
\label{s_obser}

The observational data used in this work were part of the HST's 16th 
anniversary, which was celebrated with the release of the color
composite image of M82. The data
were obtained by the Hubble Heritage Team \citep{Mut07} using the ACS 
Wide Field Channel in 2006 March, and released in fits format.
Observations consisted of 96 individual exposures in
F435W, F555W, F814W, and F658N filters covering a field of view of
$8^\prime\times8^\prime$ centered on the galaxy nucleus, and cover the entire
optical disk of the galaxy. Bias, dark, and flat-field corrections were 
carried out using the standard pipeline process (CALACS) by the Heritage Team. 
The IRAF/STSDAS Multidrizzle task was used to 
combine images for each filter and to produce weight maps, which indicate 
the background and instrumental noise. Also this task was used to identify 
bad pixels, to perform sky subtraction, cosmic ray rejection and to eliminate 
artifacts \citep[see more details in][]{Mut07}. The final data set of fits
files of science quality images have a spatial sampling of 
0.05~arcsec\,pixel$^{-1}$, which corresponds to 0.88 pc\,pixel$^{-1}$ at 
the M82's distance of 3.63 Mpc \citep{Fre94}.
In this work, we used the images in F435W, F555W, F814W bands, which we
refer to as $B$, $V$ and $I$ images, hereinafter.
The exposure times, estimated detection limits for point
and extended sources in each filter are given in Table~1.

A circle of 500~pixels (450~pc) radius centered on the starburst nucleus
is used to separate the nuclear 
region from the disk. The clusters inside this radius are associated 
with strong H$\alpha$ emitting complexes, and hence are younger
than 10~Myr \citep{Mel05}. On the other hand, the disk outside the 450~pc 
radius shows characteristic signatures of post-starburst conditions, 
with hardly any \ha\  emission. The named regions A, C, D, E and H of
M82 fall into the nuclear cluster class, whereas the B, F, G and L 
are disk clusters \citep[see ][for identification charts]{Oco78, Oco95}.
At the resolution provided by the HST, only regions H, F and L retain their
identity as bright knots, the rest being resolved into complexes of compact 
knots.

\subsection{Data extraction and source selection}

Selection of an unbiased sample of cluster candidates requires the use of
an automatic object detection code. We used SExtractor \citep{Ber96}
for this purpose. An {\it area} consisting of at least 5 adjacent background 
subtracted pixels with intensity above $k\sigma$ was defined as a 
source ($\sigma$ being the local rms value). The number of detected sources
depends critically on the value of $k$. We found $k=$5 in the $B$ 
and $V$-bands, and $k=10$ for the $I$-band as optimum values in the 
search for clusters. This condition establishes a signal-to-noise ratio greater
than 50 for a typical source of 100~pixel area.
In relatively isolated regions, the {\it area} is determined by the 
intensity profile of the source, and a pre-defined value of $k\sigma$,
whereas in crowded regions, it is determined by a deblending parameter.
Another critical parameter controlling the source detection is the 
value of the local background, which is measured using 
a {\it boxsize} of $40\times40$~pixels.  SExtractor determines Full 
Width at Half Maximum (FWHM) for each source by fitting a Gaussian profile.

We carried out independent searches of candidate sources on each of 
the $B,V,$ and $I$ images. This resulted in 44274, 82515 and 151565 sources 
in the $B,V,$ and $I$-band images, respectively. For sources identified in a
given band, we have carried out multiple aperture photometry in the other two 
bands. The list of identified sources in each band
contains both resolved (extended) and unresolved (stellar-like) objects.
The point sources have a size distribution that peaks at a FWHM of 2.1~pixels, 
with the tail of the distribution extending to 3.0~pixels,
which at the distance of M82 corresponds to a physical size of 2.6 parsec.
Hence, clusters with a Gaussian FWHM larger than 2.6 parsec are resolved 
objects, and can be easily
separated from the stars. We found that 17\% ($B$), 23\% ($V$), and 40\% ($I$) 
of the sources in the initial sample are star-like (FWHM$<3.0$~pixels). 
A visual inspection of the extended sources reveals that all the bright ones 
are symmetrical as expected for bound systems.
However, a considerable fraction 
of the fainter sources lack such symmetry, and 
the likelihood of them being clusters is rather low.
We found three kinds of contaminating sources ---
1) sources formed by improper subtraction of a local background,
2) elongated sources formed by chance superposition of several point sources,
and
3) groups of stars without a well-defined peak.
The first kind of source is due to the presence of intersecting 
dust filaments of different sizes. Contamination by this kind of sources, 
as expected, is maximal for the $B$-band, and is less important
at longer wavelengths. On the other hand, the second and third types of 
contaminations are most severe in the $I$-band image due to a large number of
red stars present. We adopted the following selection criteria to choose 
real clusters from the SExtractor source list: 

\noindent
1. All sources should have an $area\ge50$~pixels, which ensures that 
pseudo-clusters (faint superposed stars) are rejected. However, this 
criterion rejects all clusters fainter than $B=24$~mag. The most 
compact clusters (FWHM=3 pix) should be brighter than $B=21.8$~mag in order
to satisfy this condition (see Table~1). 
\\
2. The sources should obey the condition $area\ge\pi(FWHM/1.4)^2$, 
which automatically rejects all diffuse artificial sources 
created by a residual local background. It also removes faint groups of
stars on all scales. \\
3. If the {\it area} is less than 100~pixels, additionally we demand that the
sources should be nearly circular (ellipticity $\epsilon \le 0.1$), a 
condition that rejects linearly superposed double or multiple stars. 

\noindent
The second selection criterion works exceptionally well in 
discarding non-cluster sources
at the fainter magnitudes.  However, it has the disadvantage that it
rejects genuine clusters in the crowded nuclear region. This is because
the SExtractor $area$ in these regions is delimited by the 
presence of a close neighbor, 
before the intensity profile reaches 
the $k\sigma$ limit. Clusters rejected due to crowding have typical
areas less than 100~pixels. We found that these clusters distinguish 
themselves from the artificial sources by showing a relatively higher 
peak surface brightness. We used the aperture magnitude, $m_{\rm peak}$, 
within a diameter of 2 pixel as a proxy of the source's peak surface 
brightness. In order to recover these small-area nuclear clusters, we 
have established the following additional criterion:

\noindent
4. $m_{\rm peak}$ should be brighter than $m_{\rm peak}({\rm lim})$ for 
small-area clusters ($area<100$~pixels), irrespective of conditions 2 and 3. 
The values of $m_{\rm peak}({\rm lim})$ were chosen to be 25, 24 and 22 
magnitude for the $B, V$, and $I$ images, respectively.

\noindent
It is important to note that the number of nuclear clusters
selected depends critically on the value of $m_{\rm peak}({\rm lim})$.
The chosen values represent a compromise between overpopulating the sample
by un-physical clusters, and rejection of genuine clusters. 
Every condition adopted in the present work was set after an elaborate 
interactive process of visual inspection of several 
selected and rejected sources.
In Table~2, we present the statistics of the sources detected in every band.
The total number of sources identified by the SExtractor in each of the three 
bands is given in the first row of the table. The number of extended sources 
in this list is given in the second row. The third row gives the number of
candidate clusters, which is a subset of the extended sources that occupy an 
area of at least 50 pixels. The last row gives the number of clusters 
detected after applying all the selection criteria listed above. 
It can be seen that only around 6\% of the candidate clusters survive our 
selection criteria. 
From a visual inspection, we have confirmed that none of
the bright sources are rejected, and the large rejection fraction is mainly
because of over-dominance of the faint sources in the list. 
As discussed earlier in this section,
the contaminating sources at the faint end are formed by multiple stars, 
either physical systems or sources formed by chance
superposition of individual stars, which is expected given that the disk of
M82 is almost oriented along the line of sight.

A list of cluster candidates is obtained from the sum 
of the sources in the three filters, the common sources being counted once. 
%
Background extragalactic sources seen in the halo of M82 were removed 
from this list. Our final list contains 653 clusters, 260 of them belonging 
to the nuclear region. Majority of these clusters (65\%) 
was identified
in the $B$-band image. The remaining 35\% of the sources were identified 
in the $V$ and $I$-band images. Though these latter sources failed to
satisfy the $k\sigma$ criterion in the $B$-band, they have sufficiently 
good quality photometry for a reliable estimation of their masses.
The FWHM of a source is taken from the list in the shortest of the
three bands where it is detected above $k\sigma$ limit.

\citet{Mel05}, using the HST/WFPC2 images of M82, searched for nuclear SSCs 
with associated \ha\ emission, and found 197 SSCs. About 50\% of these 
clusters overlap with at least one 
of our nuclear sources. We found that the remaining sources are bright
in \ha\ with only a diffuse emission in the $B$-band, which are not considered
as genuine clusters in our selection criteria. In total, 
about 150 nuclear clusters of our sample are cataloged for the first time.

Though our final sample contains a tiny fraction of the potential
cluster candidates, the cluster size distribution function (CSF) 
obtained from clusters brighter than $B=22$~mag
is a true representation of the intrinsic function, 
as we demonstrate using Monte Carlo simulations in \S3. 
The CSF obtained from this bright cluster sample 
may not necessarily hold for the entire sample.
Future methods of disentangling multiple stars from clusters (e.g. 
analysis of 2-dimensional profiles along with colors) would be required
to investigate this issue. 
Nevertheless, given that the bright clusters in the nucleus and the disk
of M82 are also the most massive members in the respective zones, a comparison
of the derived CSF in the two zones would be invaluable in understanding the
formation mechanisms and the subsequent disruption of the clusters.
With this in mind, we carry out the analysis of the CSF for clusters 
brighter than $B=22$~mag in this work.

\subsection{Aperture photometry and luminosity functions}

With the help of SExtractor, we carried out aperture photometry in a 
number of concentric apertures for each cluster. As clusters span a 
wide range of sizes, and the background is highly non-uniform, a varying 
aperture diameter should be adopted in order to minimize the error on the 
extracted photometry. In the crowded nuclear region, special care should be 
taken in order to ensure that a given aperture does not include a 
neighboring source. We identified 
some isolated clusters and studied their growth profiles in order to 
estimate the flux fraction that would lie outside an aperture of 5 pixel 
radius, as a function of the FWHM. In general, this correction factor 
lies between 0.5--1.0 magnitude with a monotonic dependence on the FWHM, 
and showing a dispersion as large as 0.5~mag. The corrected
magnitudes agree well with the $isocor$ magnitudes obtained by SExtractor
by summing all pixels above the $k\sigma$ contour limit. Hence, we 
adopted the $isocor$ magnitudes as our photometric value. The 
colors were obtained from the photometry of the brightest part of the cluster,
corresponding to the magnitude of the fixed aperture of  5 pixel radius, 
as we do not expect color variations across the face of a cluster. Yet,
the enigmatic cluster F and its neighbor L, do show color variations of 
such kind, but those cases are exceptions rather than the rule \citep{Bas07}. 
The procedure adopted in this work 
ensures that the error on a color is significantly 
smaller than that on a magnitude. 

In Table~3, we list the results of the aperture photometry for the detected 
clusters\footnote{The printed version of the article contains the 
first 15 lines of the table, with full table available only electronically.
An extended ASCII table containing the coordinates and many other SExtractor
output parameters is available on request.}. 
Column~1 contains our identification numbers, which run from 1N
to 265N for the nuclear clusters, followed by 1D to 393D for the disk 
ones (5 nuclear clusters 84N, 139N, 147N, 149N and 258N were rejected
as the error on at least one of the colors was unreasonably high). The clusters
are arranged in increasing order of $B$ magnitude, separately for the 
nuclear and disk sources. Columns 2--5
contains the $B$ magnitude, its error, $B-V$, and $V-I$ colors, respectively.
The error column lists an estimation of the formal error, as
given by the SExtractor, errors on the colors are expected to be
of this order. The systematic errors due to an improper background subtraction,
which affects a magnitude measurement, and not a color measurement, could be 
as large as 0.2~mag for sources fainter than $B=21$ mag. 
The last column of the table contains the cross identifications with 
\citet{Mel05} (numbers following the letter M), and other named regions
mentioned in the beginning of this section.
In Figure~\ref{fig_stamps}, we present the stamp-size images of 
$50\times50$ pixel format centered on all the detected disk clusters. 
It can be seen that even the relatively faint sources have the morphology 
resembling a cluster. Identity of a few of the relatively fainter
sources as genuine clusters is debatable. However, we find 
that the contamination from non-cluster sources is less than 5\%.

\begin{figure}
\epsscale{1.30}
\figurenum{2}
\plotone{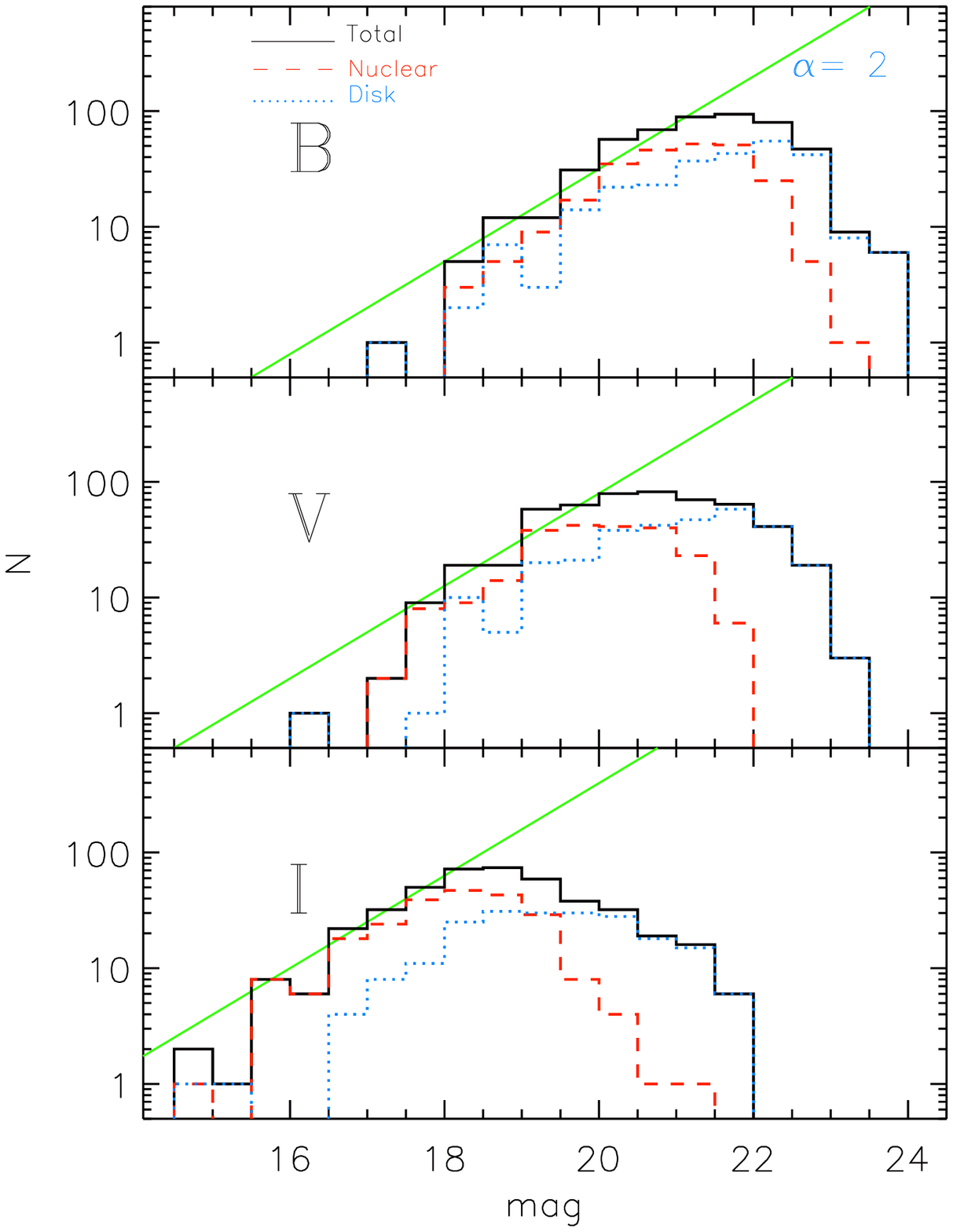}
\vspace*{-5mm}
\caption{Magnitude histogram of detected clusters in the $B,V$ and $I$-bands.
Separate histograms are drawn for the nuclear and disk clusters. The solid line
corresponds to a power-law of index of 2.0 in the luminosity function.}
\label{fig_mag_hist_ssc}
\end{figure}
In Figure~\ref{fig_mag_hist_ssc}, we show the luminosity function of sources 
satisfying our selection criteria of clusters for each of the $B,V$ 
and $I$ images, independently. The luminosity functions in every band show 
an apparent turn-over with the distributions being consistent with a 
power-law form of index $\alpha=2$ on the brighter side, and falling steeply
on the fainter side. Selection criterion 1 is responsible for the steep
fall as discussed in \S2.1. 
The turn-over magnitudes are equal 
to $\sim$21.5, 20.5, and 18.5 in the $B,V$ and $I$-bands, respectively. 
At the very bright 
end of the luminosity function, when the numbers expected from the 
extrapolation of the power-law are below 10, the observed numbers are 
systematically lower in all the bands. It seems that we are missing around 5 
bright clusters. This may suggest the existence of an upper limit to the 
cluster masses as advocated by \citet{Gie06a}. On the other hand,
it is also possible that the under-abundance of bright clusters is due to 
the combined effects of high extinction, and small number statistics.
The nuclear and disk cluster samples present similar luminosity functions, 
with the apparent turn-over magnitude $\sim0.5$~mag fainter for the latter 
sample.

\subsection{Cluster size distribution}

SExtractor provides the FWHM of a fitted Gaussian profile for each source.
We have adopted this figure of merit as a measure 
of the size of the clusters, instead of the often used half-light radius          
($R_{\rm eff}$). This is because the Gaussian FWHM is uniquely defined for
a source, and can be measured with the same accuracy for bright and faint 
regions, whereas the value of $R_{\rm eff}$ is very sensitive on the 
parameters of the fitted analytical function \citep{Lar04}. 
SExtractor calculates the half-light radius, that we found
is related to the FWHM (both in pixel units) by the equation:
 $R_{\rm eff} = 1.0 + 0.33\times {\rm FWHM}$.
The values of $R_{\rm eff}$ calculated from this relation may differ slightly
from those values determined based on an analytical function. We use this
equation only for the purpose of comparing the mean FWHM obtained in this 
work with $R_{\rm eff}$ quoted in the literature for other galaxies.

\begin{figure}
\epsscale{1.30}
\figurenum{3}
\plotone{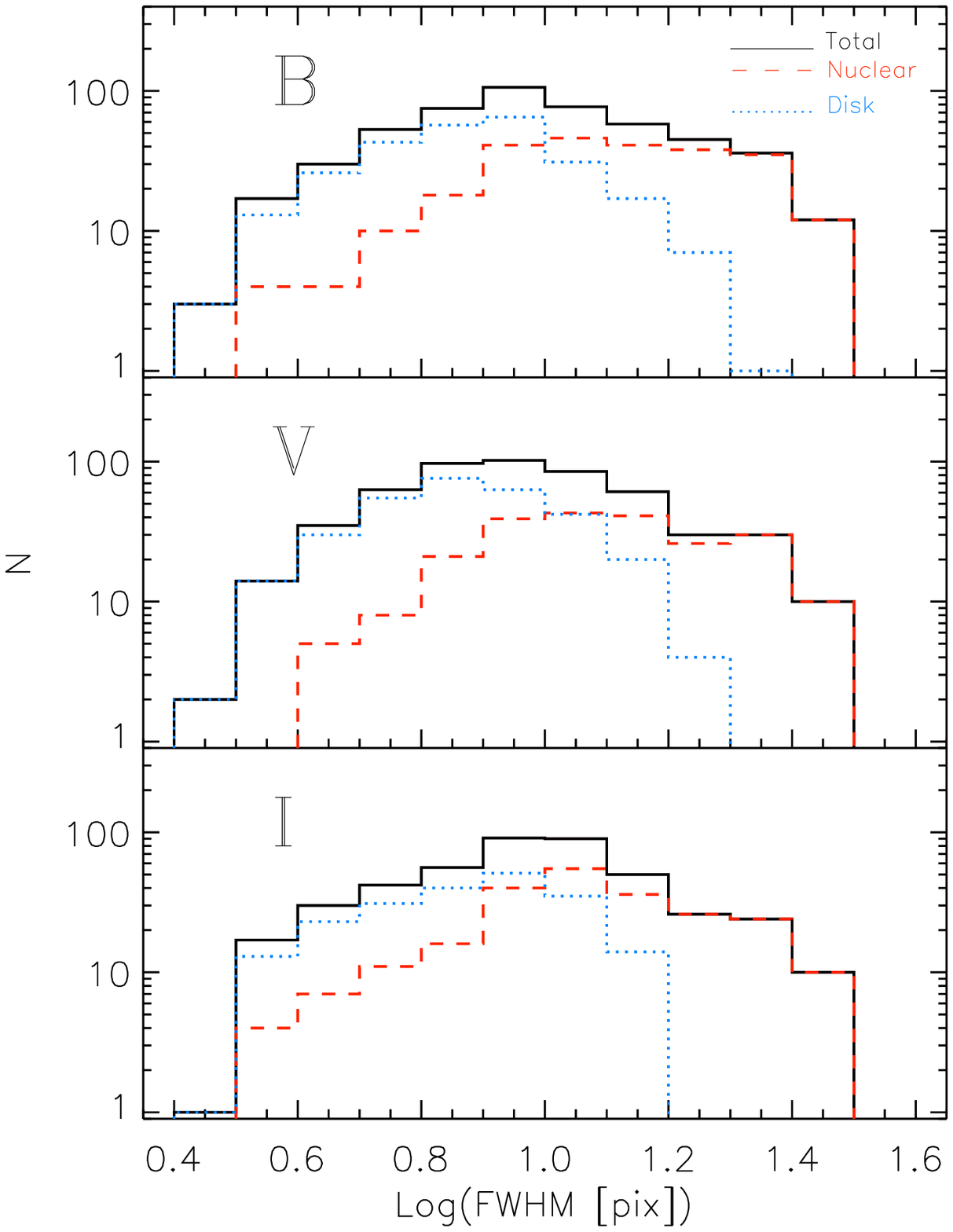}
\vspace*{-5mm}
\caption{The observed cluster size distributions are plotted for the
$B,V$ and $I$ photometric bands for the entire, the nuclear, and disk
cluster samples considered in this work. In every case, we see that the
size distribution function is log-normal with a peak at about 10 pixels.
\label{fig_fwhm_hist_all}
}
\end{figure}
The histograms of Gaussian FWHMs in the $BVI$ photometric bands, for 
the total, nuclear and disk samples, are presented in 
Figure~\ref{fig_fwhm_hist_all}. There, it is evident that the observed CSF
peaks at a characteristic value of about 10 pixels 
FWHM or $R_{\rm eff} = 3.8$~pc. \citet{Mel05} found a mean radius value 
$5.7\pm1.4$~pc for the M82 nuclear clusters. However, their values cannot be 
directly compared with ours as they had defined the radius as the inflexion 
point of the photometric growth curve in $B$ and H$\alpha$ bands.
On the other hand, our values are in good agreement with those measured in
a sample of 18 nearby spiral galaxies, where \citet{Lar04} reported a mean 
value of $R_{\rm eff}=3.94\pm0.12$~pc. 
The presence of a characteristic cluster size is quite different to what 
was found in M51, where the size distribution follows a power-law function 
\citep{Bas05, Sch07}. 


\section{Monte Carlo simulations}

The observed cluster luminosity function follows a power-law at the
bright end, turning over sharply at faint magnitudes. Similarly, the CSF 
peaks at a characteristic value of $\sim$10~pixels
FWHM. In order to investigate whether observational biases are responsible
for these turn-overs, we carried out detailed Monte Carlo simulations.

We have used simulations in order to generate a $B$-band ACS image of M82
containing N$_{\rm sim}$ clusters. The luminosity function of the simulated 
clusters is defined by a power-law distribution in the 18--25 
magnitude range: $dN/dL \propto L^{-\alpha}$. 
A power-law index $\alpha=2$ was adopted, that corresponds to a slope 
of 0.4 in the $\log(dN/dmag)$ versus magnitude plot. In the simulations, each 
cluster is assumed to be round and to have a Gaussian intensity profile 
of a given FWHM. We used two functional forms for the size distribution
of the clusters:
the first one adopting a power-law function ($dN/dS\propto S^{-\beta}$)
with an index $\beta=3.3$. In the second simulation, we adopted a log-normal 
function peaking at 10~pixels with a $\sigma$ value of 2~pixels.
We generated N$_{\rm sim}=11855$ clusters, the number of extended objects 
in the observed $B$-band source list with an area of at least 35 
pixels. 
The location of the observed sources were used as their reference positions 
of the simulated clusters. This procedure simulates properly the crowding of 
sources observed in M82. We also generated point sources as Gaussian 
profiles of FWHM fixed at 2.1~pixels, with their positions and magnitudes 
corresponding to the observed values. The precise position of every 
source (stars and clusters) is generated by randomly placing them within an 
rms of 10 pixels around that in the source list.

In order to emulate as closely as possible the observed conditions,
we degraded the noise-free simulated image by adding the observed rms
noise ($\sigma_B=0.006$ count/pixel/sec), and the SExtractor-generated 
background image of M82. The SExtractor was run on the
simulated image with the same set of parameters that were adopted for the 
observed $B$-band image. The resulting catalog was passed through the same 
selection filters described in \S~2.1  in order to obtain a catalog of 
simulated clusters.

\begin{figure}
\epsscale{1.20}
\figurenum{4}
\plotone{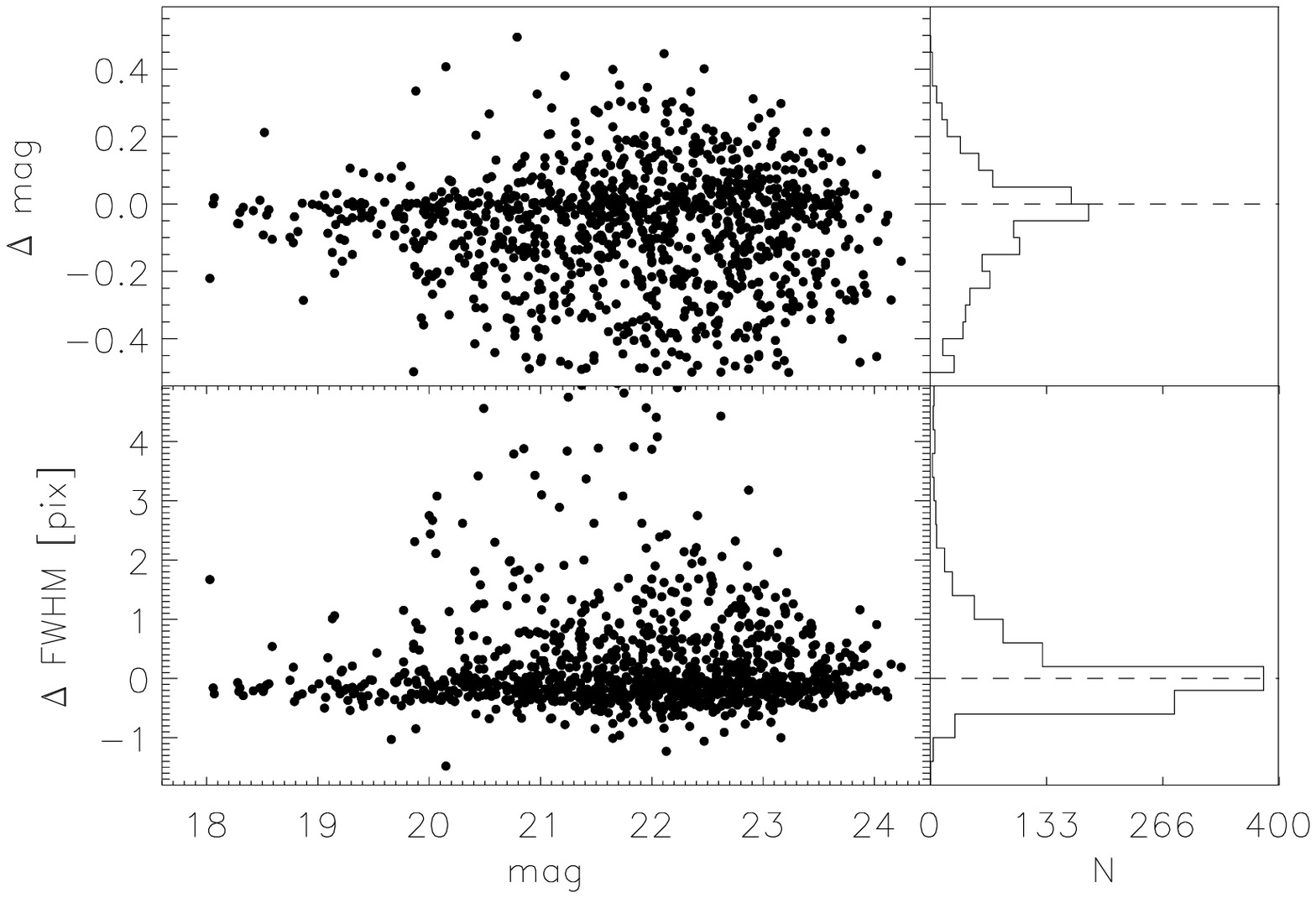}
\vspace*{-5mm}
\plotone{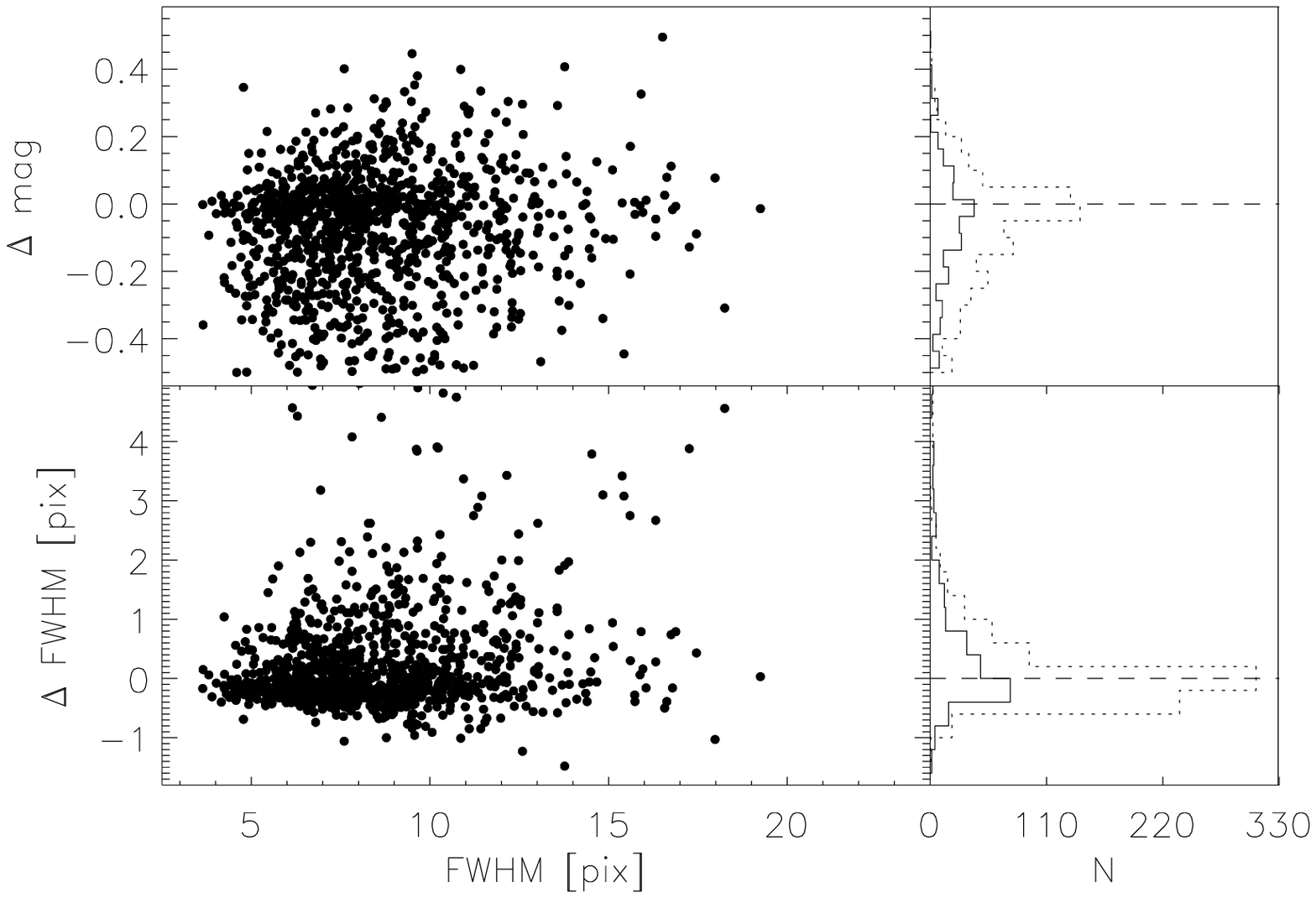}
\caption{Differences (recovered$-$original) between the
recovered B-band magnitudes and FWHMs of individual clusters
compared to their simulated values, plotted against the simulated
magnitude (top) and FWHM (bottom). The histograms of the differences
are shown for all the recovered sources in the top panel. In the
bottom panel, the histograms are shown separately for the compact (dashed line)
and loose (solid line) clusters, with FWHM=10~pixels, being the dividing value.
The results shown here are for a log-normal size distribution, but are
very similar for the power-law size distribution.
\label{fig_fwhm_mag_dif}
}
\end{figure}
We compared the extracted $isocor$ magnitude and FWHM size of every 
cluster with the simulated values. Results are summarized in 
Figure~\ref{fig_fwhm_mag_dif}. The histograms of the differences 
have a maximum at $\Delta=0$, indicating that SExtractor accurately 
recovers magnitudes and sizes of the simulated sources. For 90\% of the 
recovered sources, magnitude difference is less than 0.1~mag, while FWHM 
size difference lie within 0.75~pixel of the simulated values. For the 
remaining 10\% of the sources, the extracted magnitudes and sizes turn out 
to be systematically brighter and larger. Both effects are consequence of the
superpositions of more than one simulated source due to crowding that
extractor code identifies as a single source. Photometric recovery 
is better than 0.1~magnitude for both extended and compact sources, 
which illustrates that there is no systematic underestimation of flux,
as measured by $isocor$, from extended sources.


\subsection{Simulated Luminosity Function}

\begin{figure}
\epsscale{1.30}
\figurenum{5}
\plotone{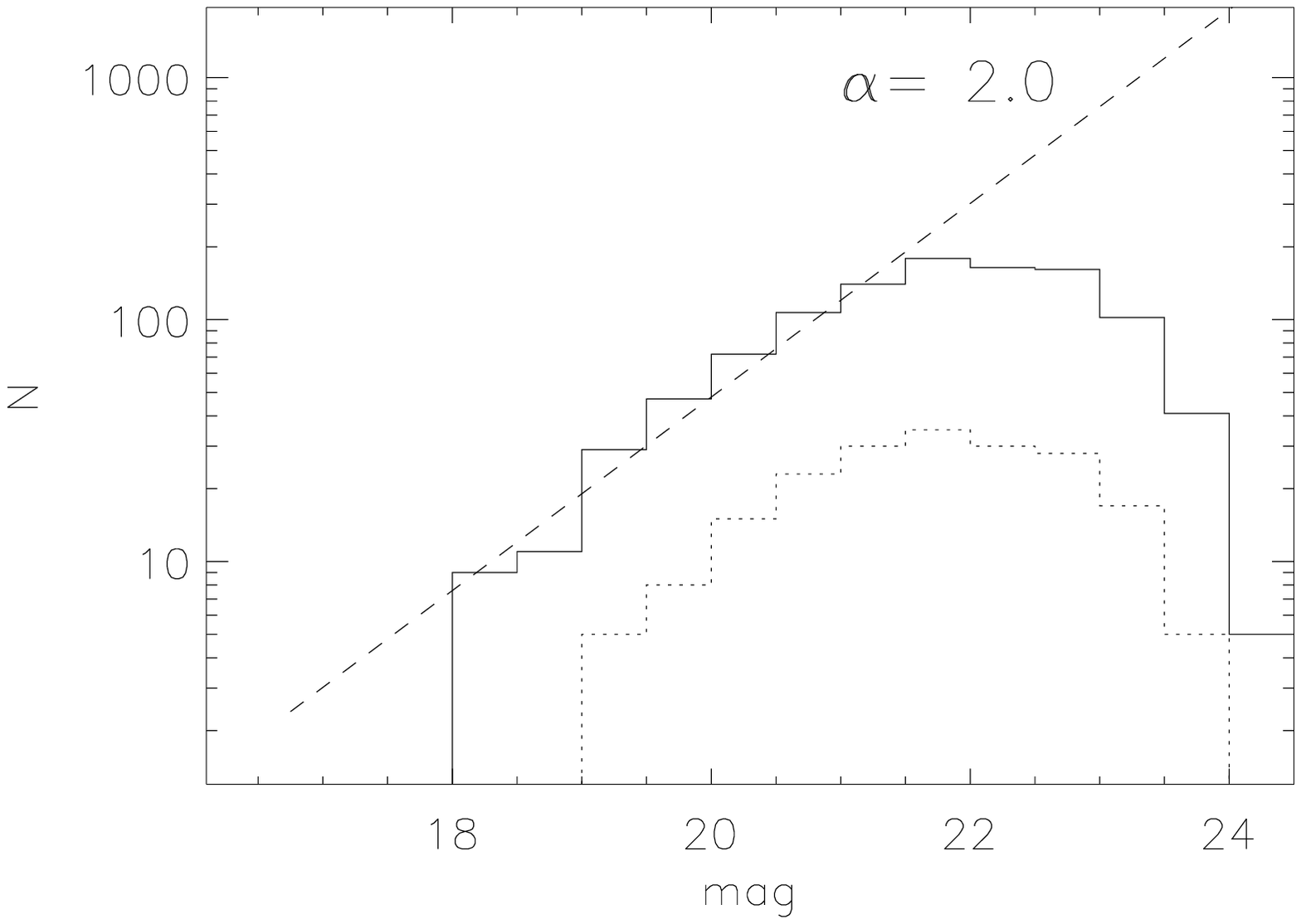}
\caption{Histogram of SExtractor recovered B-band magnitudes of simulated
clusters are shown separately for the nuclear ({\it dotted line}) and
disk ({\it solid line}) of M82. The simulated power-law function of
index=2 is shown. Recovered sources follow the power-law function
up to  $\sim21.5$~mag, beyond which the fraction of recovered sources
decreases, reaching zero at 25~magnitude.
Hence, the observed turn-over at $B\sim21.5$~mag
is an artifact of incompleteness and not a physical turnover.
\label{fig_hist_sim_mag}
}
\end{figure}
The luminosity function of the simulated clusters is plotted in 
Figure~\ref{fig_hist_sim_mag}, separate histograms are shown for the 
nuclear and disk samples. At the bright end, luminosity functions of 
the recovered sources follow the same power-law that was used to generate them.
The distribution function for the extracted disk clusters begins 
to depart from the simulated one at $B\sim21.5$ magnitude, remains nearly
flat for another 2 magnitudes before falling sharply at fainter luminosities. 
The fraction of the simulated sources recovered is larger than 90\% 
for sources brighter than 21.0~mag, falling to $\sim60$\% at 22~magnitude,
which we refer to as the apparent turn-over magnitude.
The luminosity distribution function for the nuclear clusters is very 
similar to that for the disk population except that the apparent turn-over 
magnitude is $\sim0.5$~mag brighter than that for the disk. 
The higher nuclear background, and greater crowding as compared to those 
of the disk are responsible for the difference.

The simulated luminosity function resembles very much the observed one,
implying that the observed apparent turn-over of the luminosity function 
is due to incompleteness at the faint end and not intrinsic to the cluster 
population. Hence, the intrinsic turn-over in the luminosity function, 
if any, would correspond to a magnitude fainter than $B=22$ mag.
We find that the power-law index ($\alpha=2.0$) can be recovered to within an
error of 0.1 using only the data brighter than the apparent turn-over 
magnitude, independently for the nuclear and disk clusters. This ensures that
reliable value of $\alpha$ can be determined from the observed data.
 

\subsection{Simulated Size Function}

Having established that the turn-over in the luminosity function is caused by
the incompleteness in the detection of faint clusters, we now investigate
whether the CSF is also affected by our selection criteria.
From simple analytical calculation of the type presented in Table~1, we find 
that the clusters that survive our first selection criterion all the way 
to $B=24$~mag are those with intermediate sizes (FWHM$\sim$5--11 pixels),
with the rest of the clusters having cut-off magnitudes between 22--24.
The rejection of clusters is mainly
because they fail to satisfy the requirement of at least 50 pixels area.
Thus, the CSF derived from a sample that includes clusters fainter than 
the apparent turn-over ($B=22$~mag) is not expected to represent the true 
distribution. We used the results of Monte Carlo simulations to check whether
the CSF obtained using clusters on the brighter side of the apparent turn-over
represents the true distribution.

\begin{figure}
\epsscale{1.30}
\figurenum{6}
\plotone{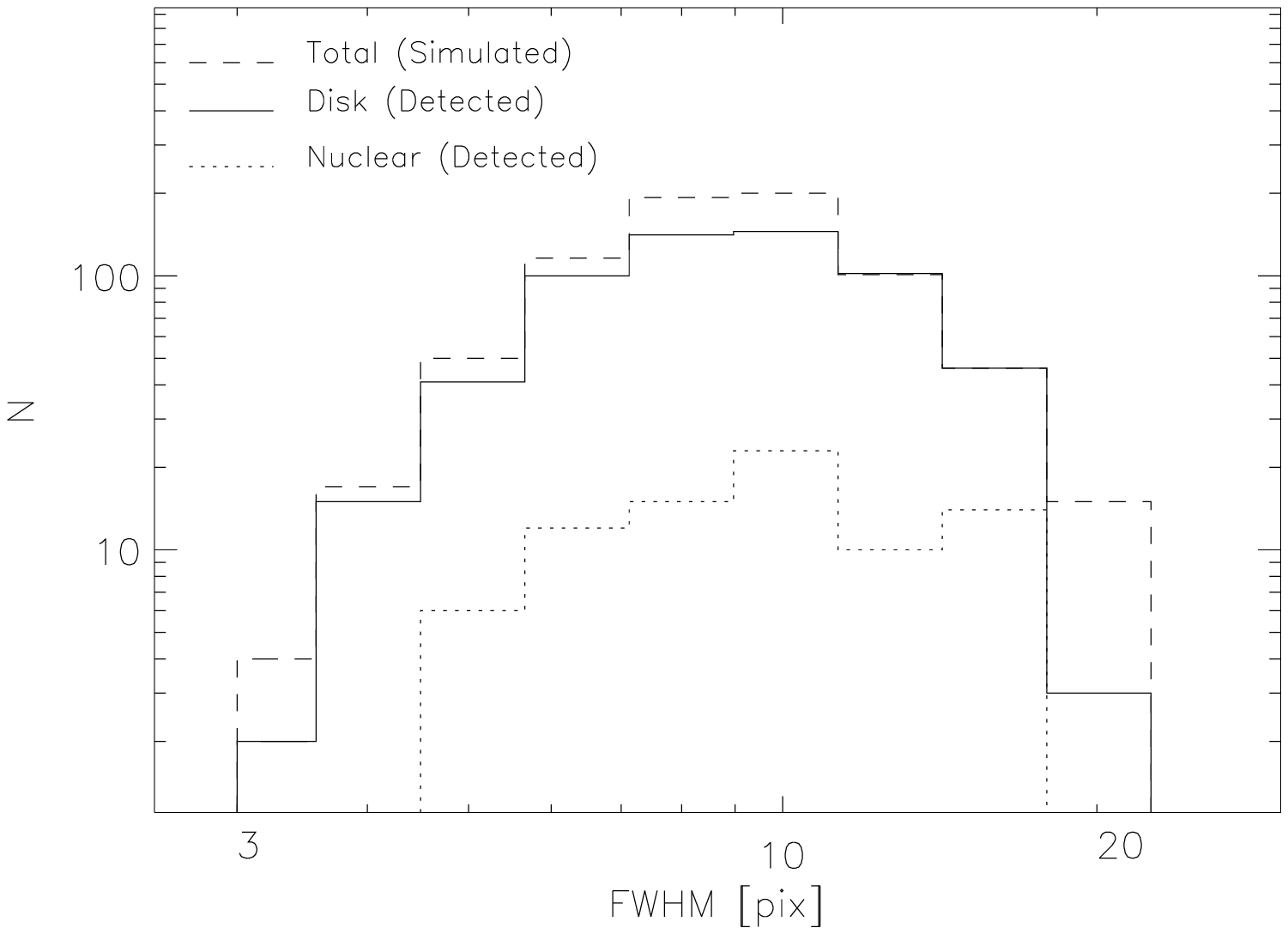}
\plotone{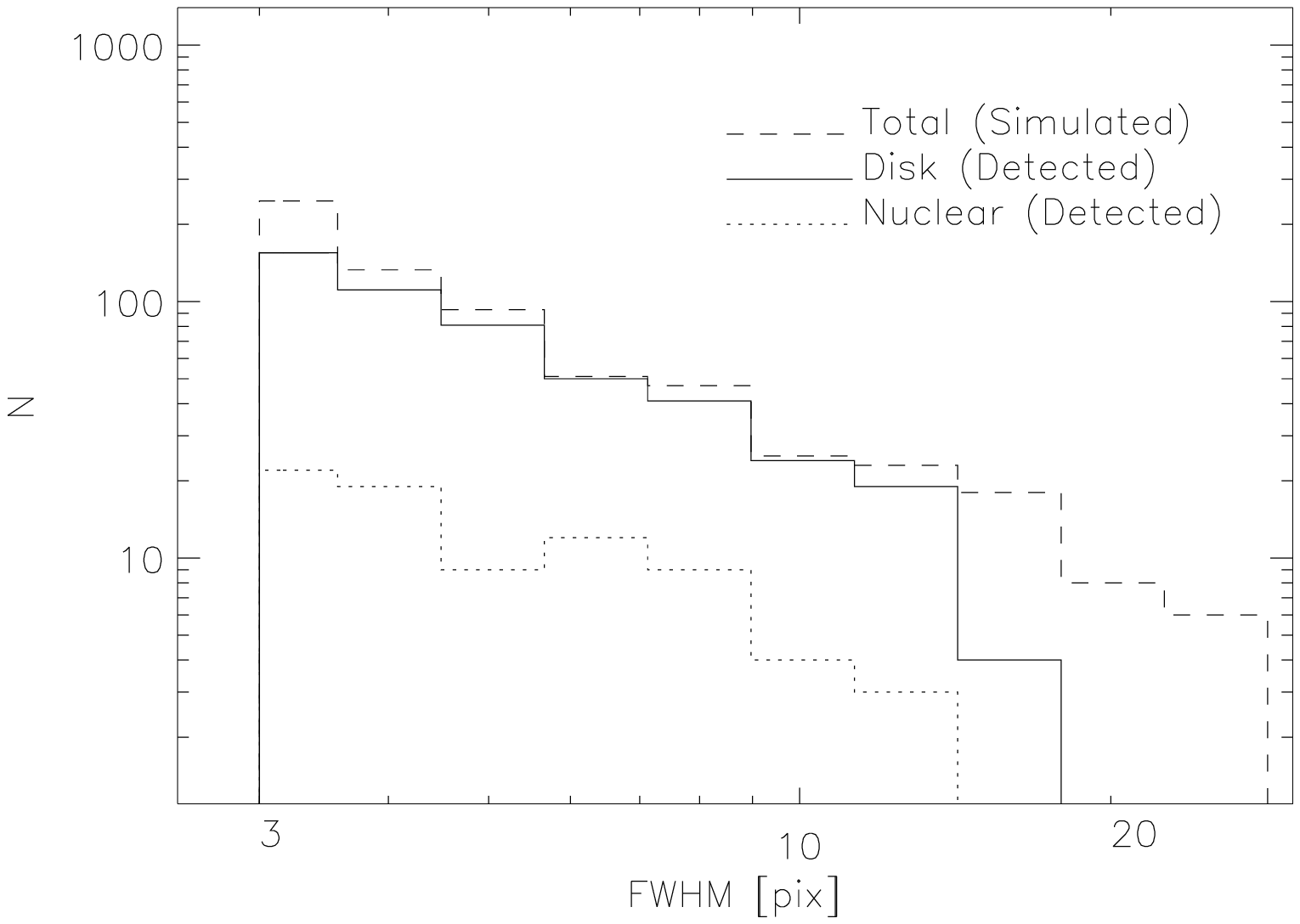}
\caption{Histograms of the distribution of FWHM of the simulated clusters
are shown for a log-normal (top) and a power-law ($\beta=3.3$, bottom)
function. Separate histograms are drawn for the clusters recovered by
the SExtractor in environments appropriately matching those of the nuclear
and disk of M82. For both the simulated functional forms, the recovered
distribution maintains its original form.
\label{fig_hist_sim_fwhm}
}
\end{figure}
We plot the size distribution function for the simulated data in 
Figure~\ref{fig_hist_sim_fwhm}. Only clusters brighter than $B=22$~mag 
in the disk, and $B=21.5$~mag in the nucleus, were used in creating these 
histograms. Separate histograms are shown for the nuclear and disk clusters.
The top and bottom panels show the histograms for a log-normal and power-law
types of distribution functions, respectively.
The recovery of the simulated size distribution functions is very good for 
both types of functional forms for FWHM$\lesssim15$~pixels. In particular, 
a power-law distribution retains its form and index value both for the
disk and nuclear samples. Thus, the observed log-normal distribution of 
sizes couldn't have been the result of selection effects transforming an
intrinsically power-law into a log-normal distribution. 
Clusters with FWHM$>15$~pixels are not detected even at
magnitudes brighter than the apparent turn-over. This explains the absence of
such clusters in the observational data (Figure~\ref{fig_fwhm_hist_all}).

From the analysis of the simulated luminosity function we have inferred that 
around 10\% of clusters brighter than $B=21$~mag, and 40\% of clusters 
brighter than $B=22$~mag are not detected. Analysis of simulated data
presented in Figure~\ref{fig_hist_sim_fwhm}
offers us an opportunity to understand the reasons for the non-detection.
One of the reasons is that they are too extended (FWHM$>15$~pixels).
Among the rest, the cluster detection fraction doesn't dependent on the value
of the FWHM. Rather, the reason for the non-detection is that either the 
cluster is situated in a region with higher than average background value 
or that there is a bright cluster in its vicinity.
Thus, our simulations illustrate that there is no size-dependent bias against 
the cluster detection as long as they are brighter than $B=22$~mag and
have FWHM$<15$~pixels. Hence,
the size function derived using data of clusters brighter than the apparent 
turn-over is a true representation of the intrinsic function.


\section{Physical Parameters of Clusters}

M82 is a spiral galaxy seen almost edge-on \citep{May05}, consequently 
the effect of obscuration is severe. Hence, the physical parameters 
derived from observational data critically depend on the treatment of the 
interstellar extinction. The issue of extinction toward the nuclear starburst 
of this galaxy 
has long been the subject of study. \citet{Rie93}, using near infrared
recombination lines integrated over the central starburst region,
estimated visual extinction values between 12 and 27
magnitudes, the exact value depending on the adopted reddening model.
\citet{For01} found $A_v\sim10$~mag in selected knots using a uniform 
foreground screen model, and $A_v>23$~mag for models where dust and stars
are mixed. In this section, we analyze the observed colors and brightness 
of the clusters with the aim of deriving extinction corrected photometry, and
compare them with a population synthesis model.

\subsection{Population Synthesis Models}

Star clusters can be considered as simple stellar populations (SSPs), 
and some of their physical parameters such as mass and age can be obtained 
from the analysis of their colors and luminosity with the help of population 
synthesis models. In this work, we have used the solar metallicity SSP 
models of \cite{Gir02}. These authors had carried out the
evolution of colors and magnitudes for the instrumental HST/ACS filters, 
a fact that enables us a direct comparison with the observed data.
The \citet{Kro01} initial mass function (IMF) in its corrected version 
has been used. It has nearly a Salpeter slope (2.30 instead of 2.35) for 
all masses higher than 1~\msun. The derived masses depend on the assumption 
of the lower cut-off mass of the IMF. In the case of standard Kroupa's IMF, 
the derived masses would be around 2.5 times higher.

\subsection{Analysis of Colors}

\begin{figure}
\figurenum{7}
\epsscale{1.35}
\hspace*{-10mm}
\plotone{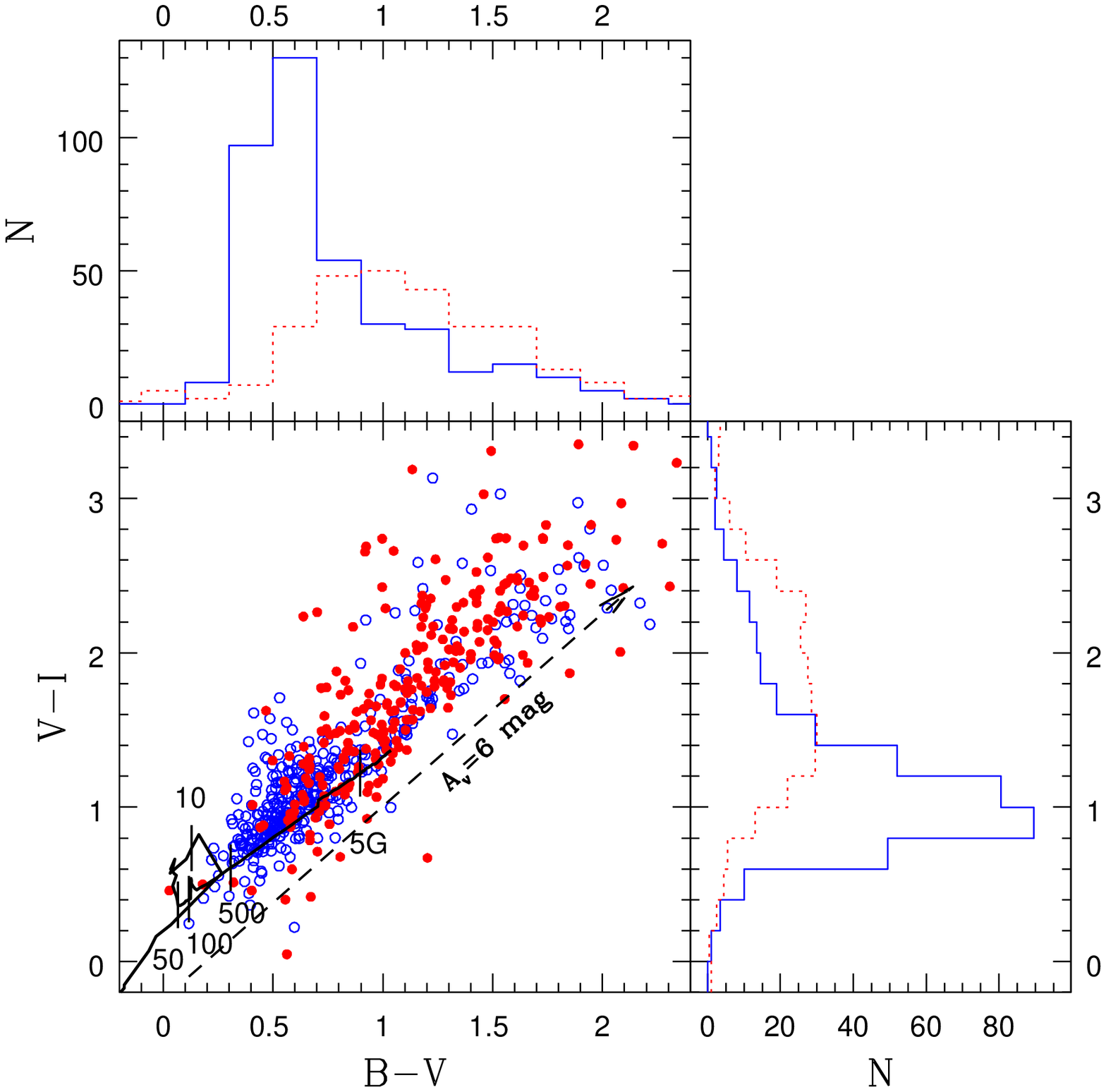}
\caption{Nuclear (solid circles) and disk (open circles) clusters in
$V-I$ vs $B-V$ plane. Evolutionary track for an SSP is shown by the
solid line, with the tick marks denoting the location at ages of
10, 50, 100, 500~Myr and 5~Gyr. The reddening vector corresponding to
$A_v=6$~mag is shown. The histograms of each axis are shown, separately
for the disk (solid line) and nuclear clusters.
\label{fig_col_vs_col}
}
\end{figure}
The observed cluster colors are plotted in the $V-I$ vs $B-V$ plane
in Figure~\ref{fig_col_vs_col}. The filled circles correspond to the
nuclear clusters, whereas open ones correspond to the disk clusters.
The solid line represents the locus of an SSP evolutionary track.
Time tags represented by vertical lines correspond to SSP ages 
10, 50, 100, 500~Myr and 5~Gyr. There, a reddening vector corresponding 
to $A_v=6$~mag is also shown. The histograms corresponding to the
distribution of the colors are shown in the top and right side panels,
separately for the disk (solid line) and the nuclear clusters.
It can be seen that the distribution of colors of disk clusters peaks 
at lower values and has a smaller spread as compared to those of the
nuclear clusters. The color distributions have a long red tail for both 
classes of clusters. 
The red tail of the distribution reaches colors even greater than those
of a 5--10~Gyr old population. Thus, the large observed color 
range is produced by a large spread in the reddening rather than a spread
in the age. Hence, the observed colors are useful in deriving the amount of
reddening, given a value for the cluster age. 


\subsection{The ages of clusters}

The stellar population in the nuclear region of M82 has been the subject of 
innumerable studies. \citet{Rie93} found that the multi-wavelength data
for the central 400~pc region of M82 can be explained by a few starburst
events over the last 30~Myr. More recent study by \citet{For03} favors 
the presence of two short-term bursts over the last 10~Myr. \citet{Sat97}
studied 12 nuclear star clusters of M82, and found their ages to be
between 4--10~Myr. \citet{Mel05} found that a great majority of the
nuclear star clusters are associated with \ha\ emission, suggesting that
they are younger than the lifetime of ionizing O stars. Based on these
results, we adopt an age of 8~Myr for the nuclear clusters. 

The \ha\ emitting regions are exclusively located 
in the central 
starburst region, and in the cone along the minor axis of M82. 
The absence of \ha\ emission in the disk implies that currently there is no 
significant amount of  
star formation. The disk is known to be rich in the Balmer 
absorption lines, suggesting an intense star formation in the 
past \citep{Oco78}. \citet{May06} made use of these spectral features to 
infer a rather young age for the disk in M82. The disk stars 
were formed in a burst as a consequence of a tidal interaction with 
members of M81 
group during the last 500~Myr. Were the clusters formed simultaneously with 
the disk-wide star formation? To answer this question, spectroscopically 
derived ages of clusters would be required.  
Such data are available only for four disk clusters{\footnote{ 
Most recently, ages for 7 more clusters have become available and lie in the
range 60--200~Myr \citep{Kon07}.}}
 --- regions F and L, and two clusters in the region B \citep{Smi06}. 
Derived ages range between 50--65~Myr for F and L, and $\sim350$~Myr for
the two optically bright knots in region $B$. We have used the ACS 
photometric data to check whether the ages of these bright clusters are 
representative of that of the bulk of the population. As discussed 
previously, the colors in M82 are strongly affected by reddening, and 
hence any age inference based on colors of individual clusters is not 
expected to give a unique solution. We investigated whether the color 
differences between the clusters and the surrounding disk could be used to 
derive a statistical age for the cluster population in the disk.
If the clusters have a similar age as that of the stellar disk, and if the 
reddening is caused by the dust clouds external to the clusters, then the 
color distribution of the clusters would be similar to those of the local 
disk. We carried out experiments to test this idea. 

\begin{figure}
\figurenum{8}
\epsscale{1.35}
\hspace*{-10mm}
\plotone{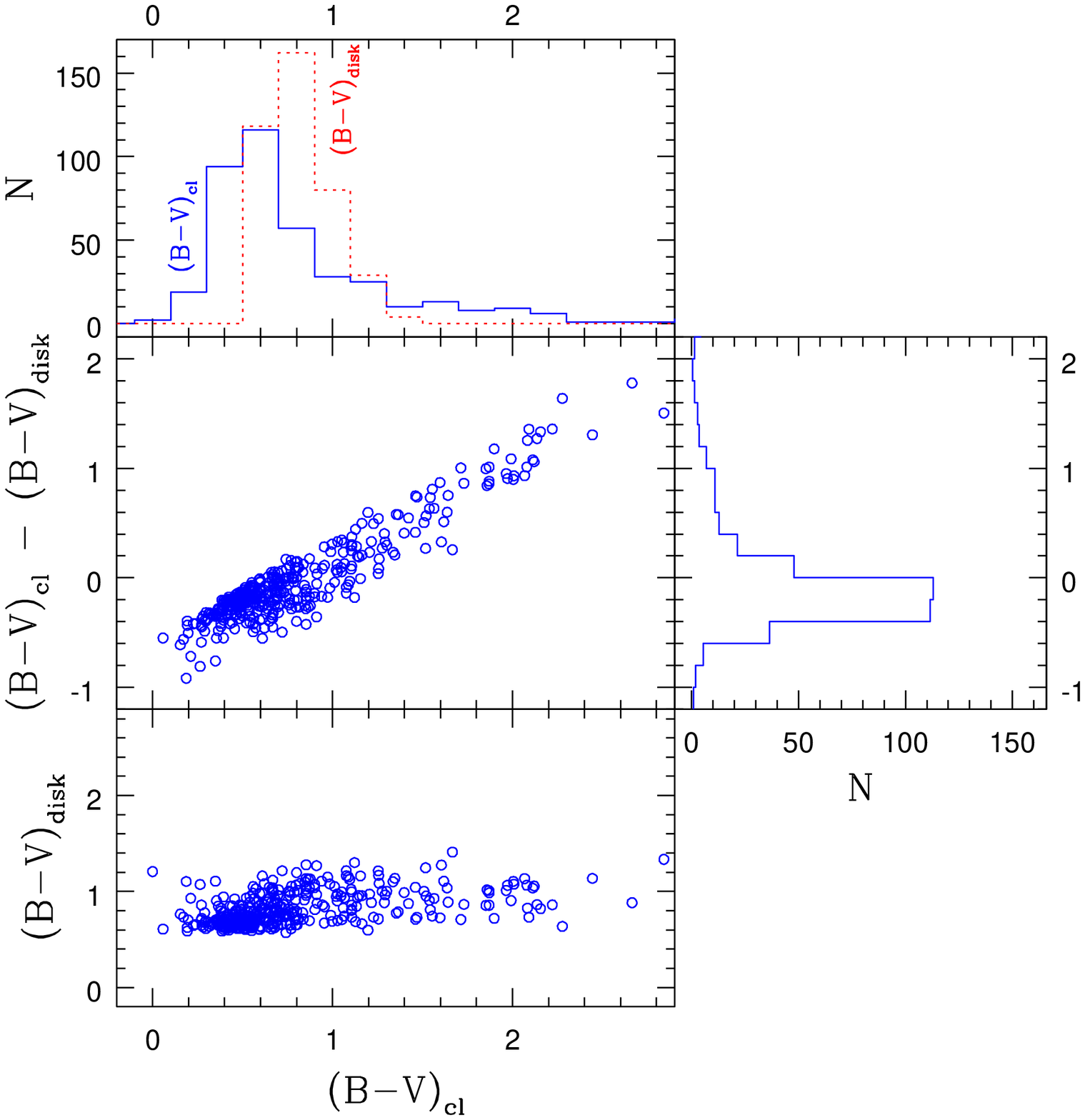}
\caption{Distribution of the $B-V$ colors of clusters compared to that of the
surrounding local disk. ({\it bottom panel}) Color of the local disk
surrounding a cluster is plotted against the color of the cluster.
({\it middle left panel}) Difference of the colors of disk clusters with
respect to that of the surrounding disk plotted against the cluster colors.
The histograms of each axis are shown. The majority of the clusters (85\%)
are bluer than the surrounding disk, suggesting that clusters in general are
younger than the disk. However, the red colors of the remaining clusters
are correlated with its ``excess'' from the disk, implying that they are
heavily reddened.
\label{fig_col_disk}
}
\end{figure}
We measured the colors of the disk surrounding every cluster in 
a 10-pixel wide annulus with an inner radii of 20, 30 and 40 pixels, that
correspond to 1, 1.5, and 2$^{\prime\prime}$, respectively. We found that
the colors in these three annuli are very similar. 
In Figure~\ref{fig_col_disk}, we compare the colors of
the inner most annulus with that of the corresponding cluster.
In the bottom panel, the disk $B-V$ color is plotted against that of the 
cluster. There, we can see that the disk colors span a much shorter range
as compared with that of the clusters.
In fact, the color difference (middle panel) of the cluster and the local
disk reflects the color of the cluster. The peak of the cluster color 
distribution is at $\sim0.2$~mag bluer than the surrounding disk.
Systematically blue colors of the clusters suggest that the clusters 
are younger than the disk. Interestingly enough, the disk surrounding the red
clusters is not as red, suggesting that the dust causing the reddening of 
the clusters is local to the cluster, perhaps associated with the 
interstellar medium left over from the star formation episode.
Very similar results are found from the analysis of the $V-I$ colors. 

The bluer colors of the clusters could be due to metallicity effects,
if the clusters have systematically lower metallicity content as compared 
to the local disk. This case is unlikely, yet it could happen if the 
clusters were formed from the accreted metal-poor gas, after the formation 
of the disk. In this scenario, clusters would be at the most as old as the 
disk. On the other hand, if the cluster stars are of higher metal abundance, 
they could be as young as 50~Myr. This rather young age compares very well 
with the 50--65~Myr determined by \citet{Smi06} for the bright clusters $F$ 
and $L$ from spectroscopic indices.  Consequently, the ages of the disk 
clusters are most likely lie in the 50--500~Myr range. For the 
determination of reddening and mass of individual clusters, we fix their 
age at 100~Myr, 
and discuss the error on these quantities if they were as 
young as 50~Myr or as old as 500~Myr.


\subsection{Color-Magnitude Diagrams}

\begin{figure*}
\figurenum{9}
\epsscale{1.25}
\vspace*{-10mm}
\hspace*{-10mm}
\plotone{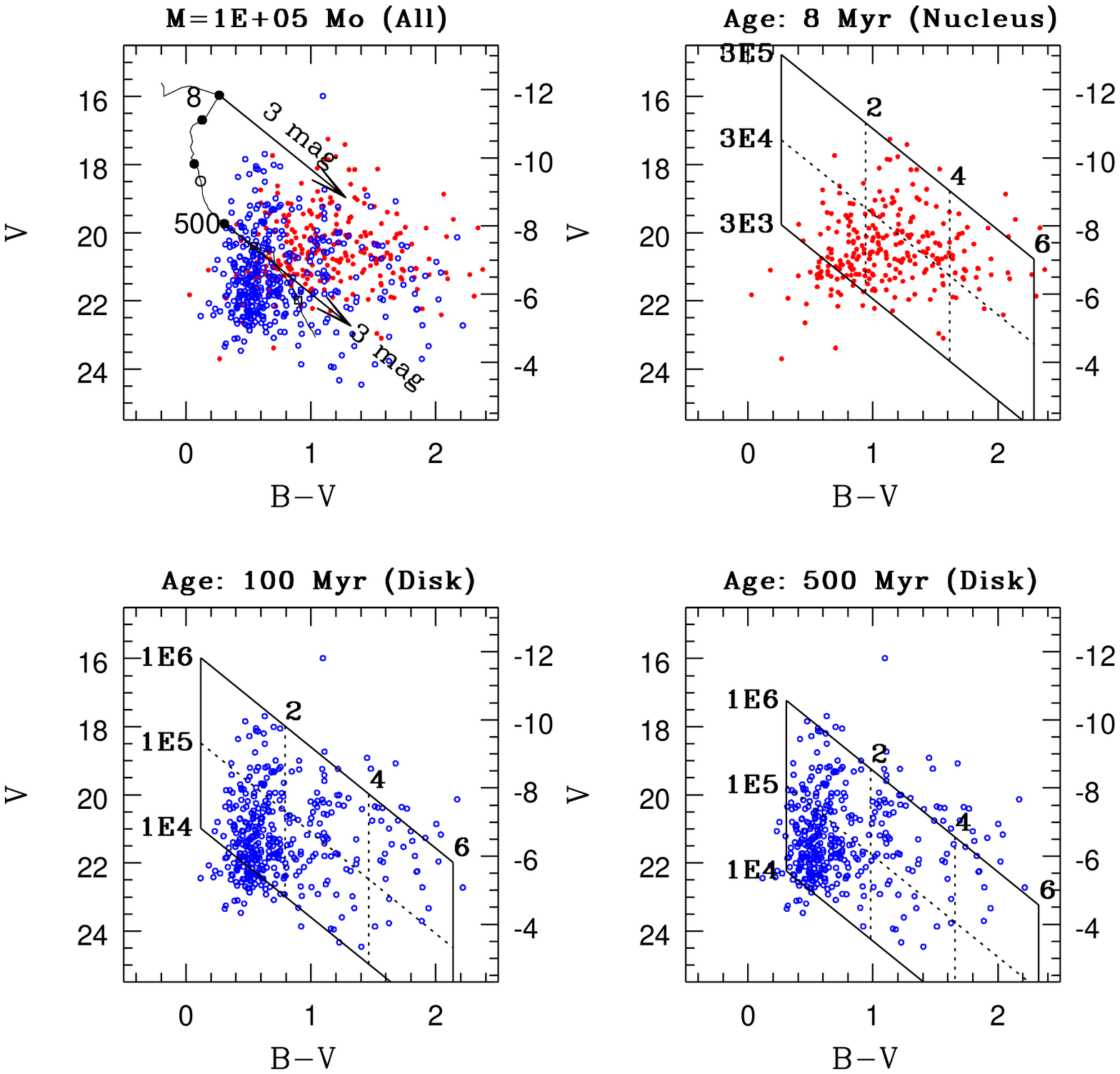}
\caption{Observed color-magnitude diagrams (CMDs) for the nuclear (filled
circles) and disk clusters, in M82. ({\it Top left}) Evolutionary track
for an SSP of a cluster mass of $10^5$~$M\odot$ is superposed.
Two vectors, placed at 8~Myr and 500~Myr, show the location of the
track reddened by $A_v=3$~mag.
In the top-right panel, we show the CMD for the nuclear clusters only.
The locations of an 8~Myr SSP for a range of cluster masses and visual
extinctions are shown by the superposed grid. Mass varies vertically
along the grid (in solar units), whereas the visual extinction (in magnitude)
varies along the diagonal axis.
In the bottom panels, we show a similar diagram for the disk clusters,
with the superposed grids corresponding to fixed ages of 100~Myr (left)
and 500~Myr (right). In all the panels, tick mark values of the
right-vertical axis correspond to the absolute magnitude.
In the derivation of mass and visual extinction values, we adopt a uniform
age of 8~Myr for the nuclear, and 100~Myr for the disk clusters.
}
\label{fig_cmds}
\end{figure*}
Once the reddening is derived using the colors, we obtain masses of individual 
clusters with the help of the SSP for the assumed age (8~Myr for the nuclear 
clusters, and 100~Myr for the disk clusters).
The method we have followed is illustrated in Figure~\ref{fig_cmds}. 
For a given position in the Color Magnitude Diagram (CMD), we derived the 
reddening by comparing the observed colors with those of the SSP. 
\citet{Car89} extinction curve with Rv=3.1 is used to convert reddening to
visual extinction $A_v$. 
The mass is then calculated using the extinction-corrected $B$-band
luminosity and the mass-to-light ratio of the SSP for the assumed age. 


In Table~3, we list the derived quantities (visual extinction, extinction
corrected B-band absolute magnitude and mass) along with their estimated 
errors for every cluster. We have independently derived extinction values 
from $B-V$ and $V-I$ colors, and the tabulated values are the mean 
of the two measurements. The dispersion from the mean value is tabulated
as the error on $A_v$. The absolute $B$ magnitude listed is 
extinction-corrected. The tabulated mass is photometric and is obtained by
the method described above, using an age of 8~Myr for the
nuclear clusters, and 100~Myr for the disk clusters. 
The derived $A_v$ would be systematically higher by 0.12~mag if the disk
clusters are as young as 50~Myr, and lower by 0.58~mag if they are as old 
as 500~Myr for the SSPs used in this work \citep{Gir02}. 
The derived disk masses would also be different if we had used a different 
age, as both the extinction and mass-light-ratio are sensitive to the assumed  
age of the population. 
Fortunately, these two corrections act in opposite directions, with the net 
result that the masses would always be within a factor of 1.5 of the 
tabulated values for the range of ages between 50--500~Myr.
The errors on $B$ magnitude and masses are calculated by quadratically summing 
the errors on $A_v$ and the $B$ magnitude, the latter taken as 0.2~mag for all 
clusters.

\begin{figure}
\figurenum{10}
\epsscale{1.20}
\plotone{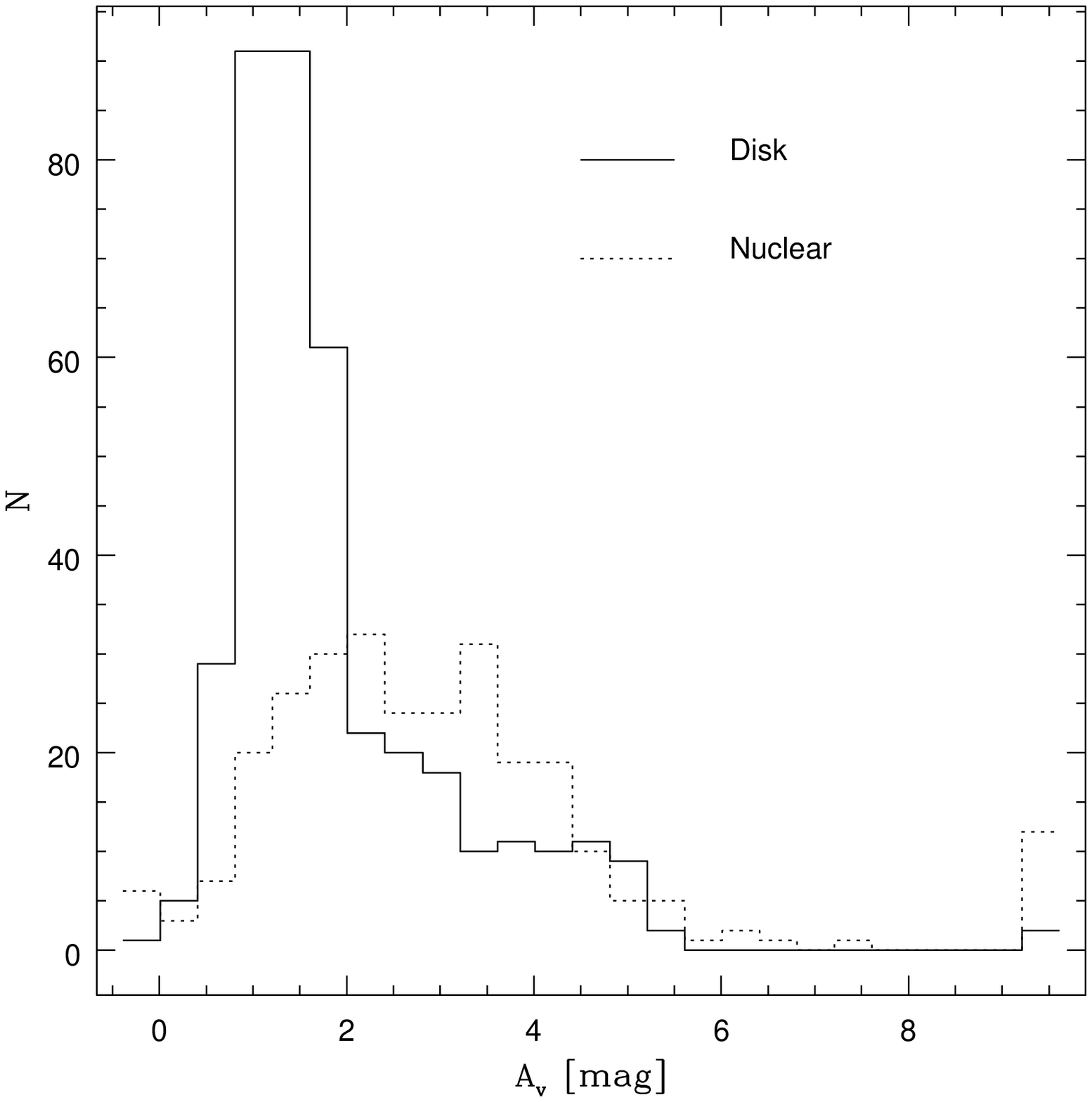}
\caption{Histogram of visual extinction ($A_v$) for the nuclear and disk
clusters. Nuclear clusters experience a higher extinction as compared
to the disk clusters.
\label{fig_physpar}
}
\end{figure}
The histogram showing the distribution of visual extinction for an 
assumed age
of 8~Myr for the nuclear, and 100~Myr for the disk clusters, is plotted in 
Figure~\ref{fig_physpar}. The distribution for the nuclear clusters is 
nearly flat between $A_v=$1.0--4.0~mag, whereas it is peaked at $\sim1$~mag 
for the disk clusters. There are a few clusters in both the disk and the 
nucleus with inferred $A_v>$6~mag. The derived $A_v$ is less than 
2.0~mag for 85\% of the disk clusters, whereas only 20\% of the nuclear 
clusters are below this value. 
The results remain practically the same if all the disk clusters are as young
as 50~Myr. On the other hand, if all the disk clusters are as old as 500~Myr, 
the distribution of their $A_v$ would peak 
at a value lower by 0.58~mag ($\approx1$ bin width) with respect to that in
Figure~\ref{fig_physpar}. This would further increase the mean difference in 
the extinction values between the disk and nuclear clusters.

\subsection{Cluster Mass Distribution Function}

\begin{figure}
\figurenum{11}
\epsscale{1.25}
\plotone{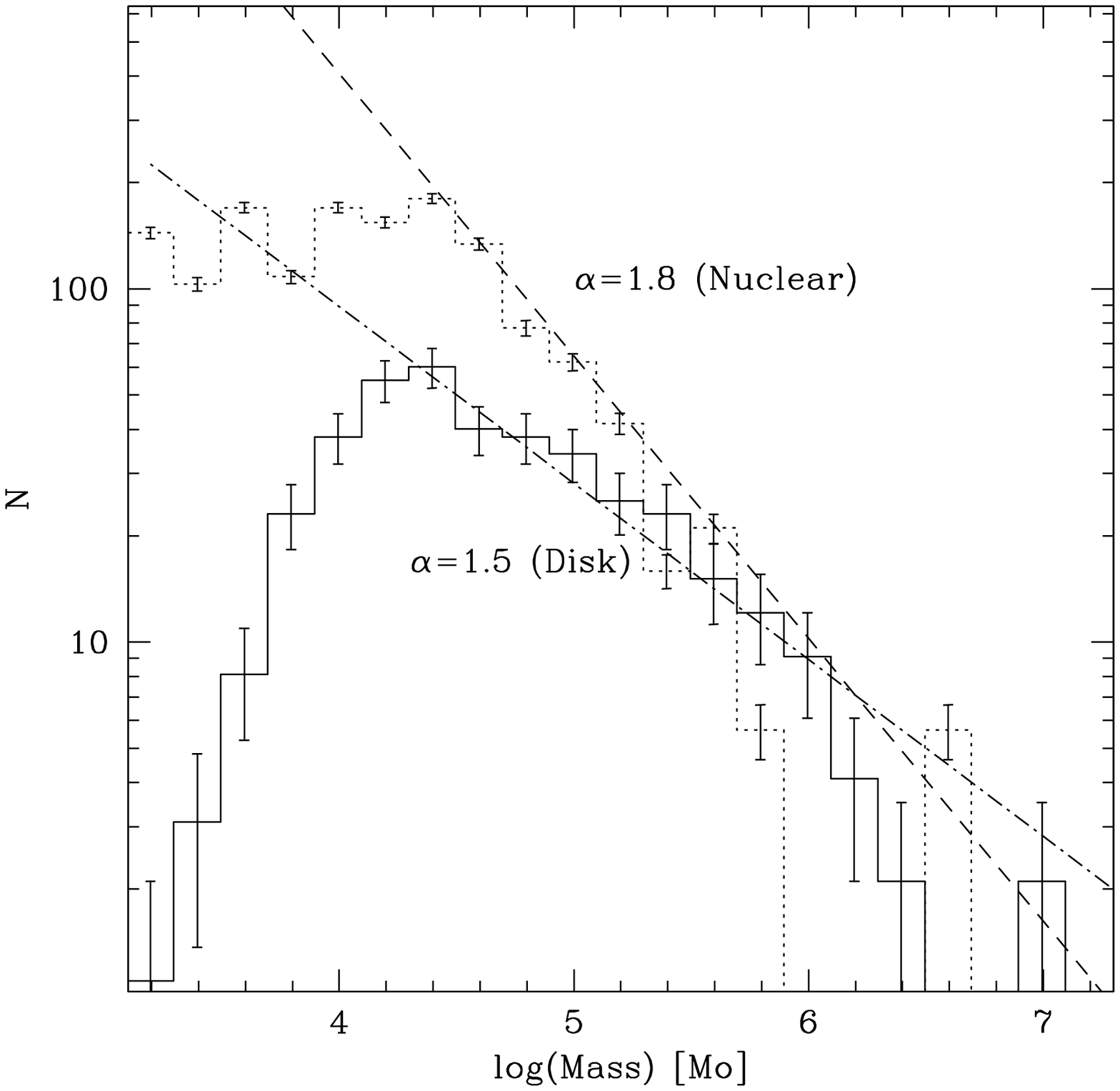}
\caption{Mass functions for the nuclear (dotted line) and disk (solid line)
cluster samples. Both the samples follow a power-law distribution between
$2\times10^4$\msun\ and $10^6$\msun. The best-fit indices in this mass
range are indicated.
\label{fig_mass_fun}
}
\end{figure}
The derivation of the cluster masses for our entire  
sample enables us to 
derive the present-day CMF. In Figure~\ref{fig_mass_fun}, we plot the CMF 
separately for the nuclear and disk clusters. The nuclear CMF is scaled up
to match the disk CMF at $1.5\times10^6$~\msun. Poissonian error bars
are indicated. The distribution for both samples follows a power-law over 
almost two orders of magnitude in mass for cluster masses 
above $\sim2\times10^4$~\msun. However, the power-law index for the disk and 
nuclear cluster populations shows a marginal 
difference, $\alpha=1.8\pm0.1$ for 
the nuclear clusters, and $\sim1.5\pm0.1$ for the disk population.
A different choice of age (within the range 4--10~Myr for nuclear clusters 
and 50--500~Myr for the disk clusters) in the derivation of the photometric 
mass would have only changed the masses by less than the size of the bin width 
used in Figure~\ref{fig_mass_fun}. Hence, the slopes of the mass functions 
are not affected by our assumption that all the clusters in each zone have the
same age. 

Studies of young star clusters in nearby galaxies yield a value for $\alpha$
close to 2.0 \citep{deG03c}. Our derived value of $\alpha=1.8\pm0.1$ for the 
M82 nuclear clusters is marginally flatter than  
this. On the other hand, the mass function of the disk clusters
is clearly flatter than that for young clusters, implying that 
the low-mass clusters 
are selectively lost as the cluster population evolves. 
In order to shed more light on the missing population, we
analyze the cluster size distribution below.


\subsection{Cluster Size Distribution Function}

\begin{figure}
\figurenum{12}
\epsscale{1.7}
\hspace*{-15mm}
\plotone{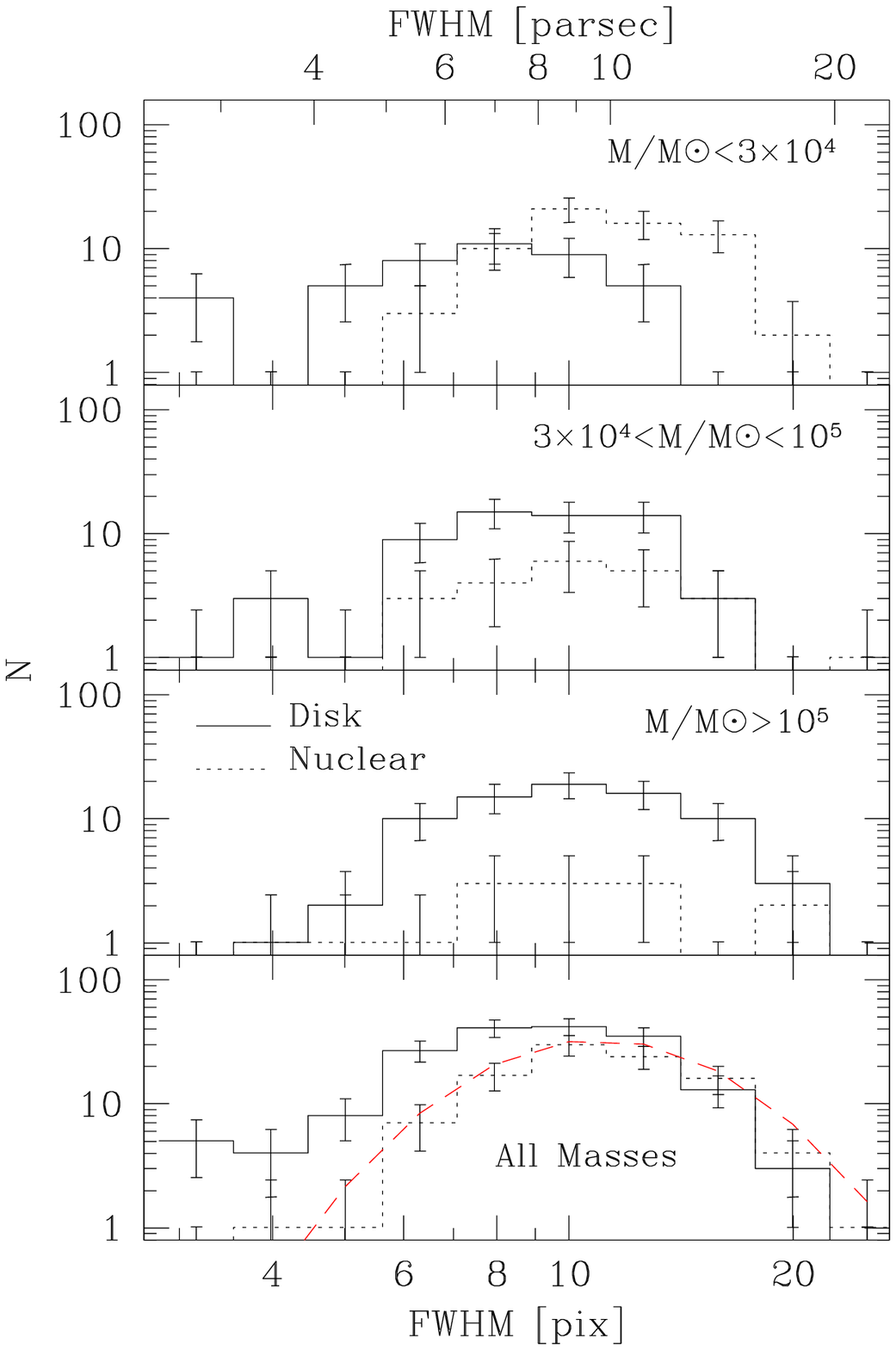}
\caption{Size function for all the detected clusters (bottom panel). In the
top three panels, the size function is shown for three different mass bins.
In each panel, histograms are drawn separately for the nuclear and disk
clusters. A log-normal distribution function is drawn fitting the
nuclear cluster sample (dashed line) in the bottom panel.
\label{fig_size_fun}
}
\end{figure}
\begin{figure}
\figurenum{13}
\epsscale{1.2}
\hspace*{-15mm}
\plotone{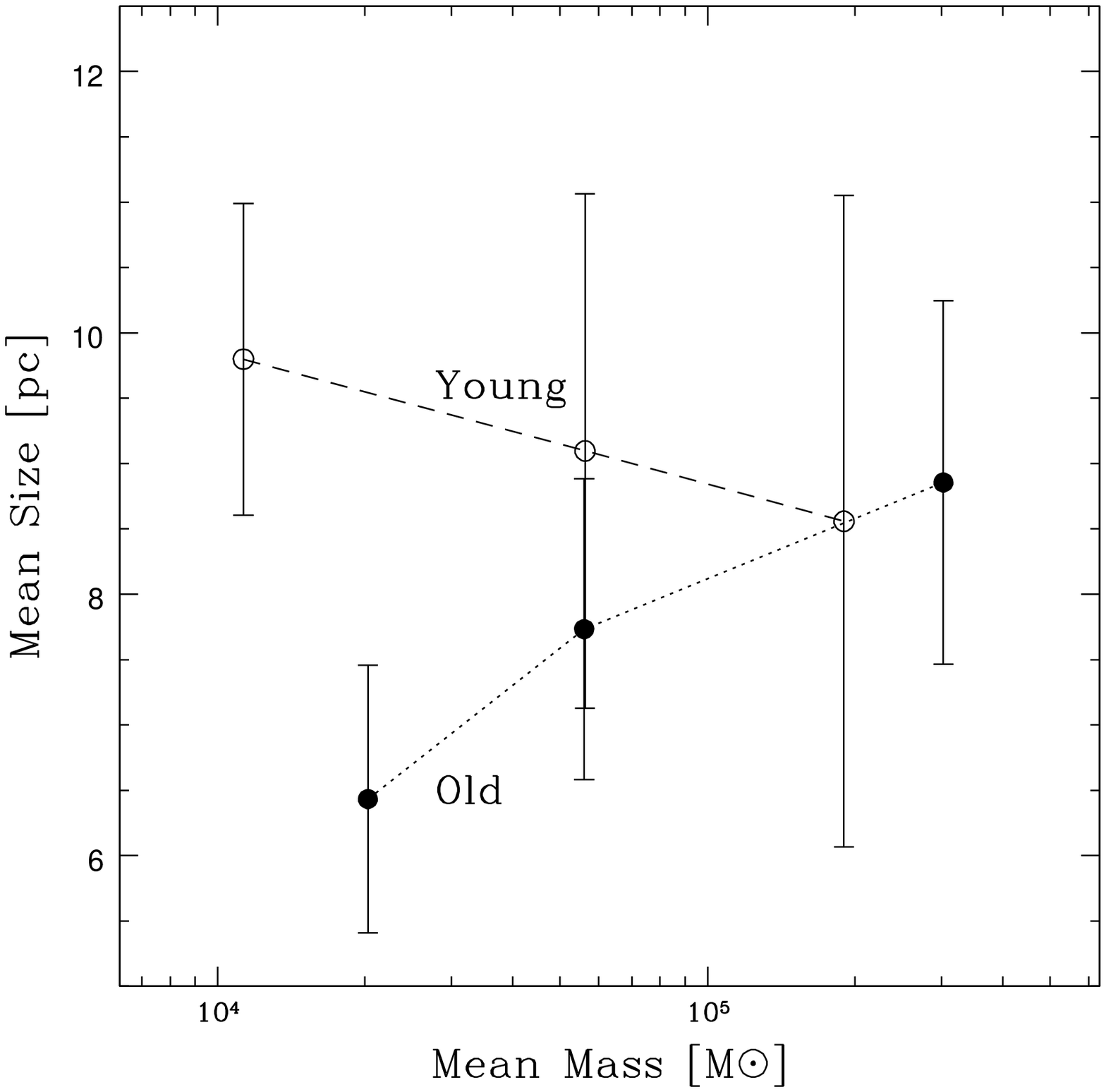}
\caption{Mean size (FWHM) of the clusters as a function of mean mass for
three mass bins for the nuclear (young) and disk (old) samples.
The error bar denotes the rms dispersion about the mean value.
High mass clusters have similar mean sizes irrespective of their
evolutionary status. On the other hand, mean size of the low-mass clusters
decreases as they become older. Among the young clusters, low-mass
ones are more extended than higher mass ones.
\label{fig_stats}
}
\end{figure}

In Figure~\ref{fig_size_fun}, we compare the 
CSF for the nuclear and disk samples. 
Following the discussion in \S3.2, only 
clusters brighter than the apparent turn-over luminosity ($B=22$~mag 
for the disk and $B=21.5$~mag for the nucleus) were considered.
The mass that corresponds to the apparent turn-over is 
$\sim2\times10^4$\msun, and hence the plotted functions are representative 
of the cluster population more massive than this limit.
The CSF for the nuclear region can be fitted very well with a 
log-normal function centered at FWHM of 11~pixels ($\sim10$~pc). 
On the other hand, the CSF for the disk clusters is quasi log-normal: 
a log-normal fit to the distribution of clusters with sizes larger than
11~pixels underestimates the actual number of observed compact clusters,
while a fit to the distribution of compact clusters shows a 
deficiency of extended or loose clusters. In order to investigate the 
possible origin of the skewness in the distribution, we have analyzed 
the size distribution function in three separate mass bins, i.e.
low-mass (mass$<3\times10^4$\msun), high-mass (mass$>10^5$\msun) and 
intermediate mass (all in between).
The CSF for the nuclear clusters follows the log-normal distribution
in each of the three mass bins (taking into account the statistical
errors caused by the small number of clusters in the highest mass bin). 
The disk clusters in the highest mass bin, also follow a log-normal 
distribution. The CSF of the lower mass clusters of the disk, however, 
depart from that of the nuclear clusters: the largest clusters, as well 
as the mean cluster size, are systematically smaller for the lower mass bins. 
This tendency is illustrated in Figure~\ref{fig_stats}, where the mean 
cluster size for each mass bin has been plotted against the mean mass of 
clusters in that bin, for the young and old ones, separately. For the highest 
mass bin, the mean sizes of the young and old clusters are similar. The 
mean size decreases systematically with decreasing cluster mass for the 
old clusters, whereas the inverse is true for the young clusters. 

Systems in Virial equilibrium are expected to follow a
power-law radius-mass relation ($R\propto M^{\gamma}$) with a $\gamma=0.5$,
the most common examples being the Giant Molecular Clouds (GMCs) and 
elliptical galaxies.
Stars of a cluster are well mixed and reach Virial equilibrium after just
a few crossing times \citep{Kin81}. M82 clusters are around ten crossing times
old and hence are expected to be in Virial equilibrium \citep{McC07}.
The relationship seen in Figure~\ref{fig_stats} corresponds to
a power-law index of $\gamma=0.05$, and $-0.02$, for the old and young 
samples, respectively, i.e. both values are significantly flatter as compared 
to those of bound systems in Virial equilibrium.
On the other hand, our values for the disk clusters are in excellent agreement
with that found ($\gamma=0.08\pm0.03$) for star clusters in nearby galaxies by 
\citet{Lar04}.
\citet{McC07} have obtained an index of $\gamma=-1$ for clusters losing 
mass adiabatically over time. Thus, mass-loss from clusters could yield 
a flatter radius-mass relationship. The similar observed range of sizes for 
high-mass clusters independent of whether they are young or old 
suggests that these clusters have reached their equilibrium sizes and
that the evolution hasn't played a role in changing their sizes.
The observed radius-mass relationship then implies that the sizes of
lower mass clusters are well above their equilibrium values, or in other
words they are distended ones. On an average, younger clusters are
farther from equilibrium values as compared to older ones.
Together, these results imply that many of the young low-mass clusters in our 
sample are loosely bound or unbound, or probably are expanding OB 
stellar associations. The extended low-mass clusters are being selectively 
destroyed as a function of time, whereas compact clusters of all masses 
have survived. However, the sizes of even the most compact 
low-mass (mass$\lesssim5\times10^4$~\msun) clusters are larger than their 
Virial expected values, implying that they are expanding and may not 
survive over the Hubble time.

The observed differences in the CSF for young and older clusters are
consistent with the expected evolutionary effects. Both, the disruption
of the loose OB associations and the dynamical trend towards relaxation
would diminish the number of large low-mass systems. Thus, the destruction 
process is both mass and size dependent, with the most extended clusters 
in each mass bin being the first ones to be dispersed. 
The near equality of the mean size for young and old clusters of masses
higher than $\sim10^5$~\msun\ implies that all high mass clusters 
survive for ages more than $\sim10^8$~yr or that the destruction process 
is independent of the size for these high mass clusters.
In the next section, we will discuss these observational 
results in the context of results obtained by other studies, and propose 
the most likely scenario of cluster disruption in M82.


\section{Discussions and conclusions}
\label{s_concl}

The presence of two populations of star clusters in M82, one 
young (age$<10^7$~yr), and the other relatively old (age$\sim10^8$~yr) 
allows us to interpret the observed differences of their distribution functions
in terms of evolutionary effects. We start this section by comparing our 
CMF for the entire disk, with that published for the region B in M82. 

\subsection{M82 B and the Initial Cluster Mass Function}

The bright blue region known as M82 B about a kiloparsec distance to the 
northeast of the nucleus has been the subject of study in a series of articles
\citep[][b]{deG01,deG03a}. These authors were able to detect more than
200 cluster candidates using the HST/WFPC2 images, which were used to
construct luminosity and mass functions. They found that the mass function
is log-normal with a turn-over in the mass function at 
 $\log$(mass)$= 5.1\pm0.1$, well above their completeness limit for 
cluster detection. Adopting an estimated age of 1~Gyr for this system, 
they argued that the log-normal mass distribution cannot be the result
of selective destruction of smaller clusters as the clusters evolve,
and instead should have been born that way.
In essence, they favored a log-normal initial CMF for M82 clusters, 
and proposed that the GCs probably were also born with log-normal initial CMF.

A log-normal initial CMF is not consistent with our dataset for younger
clusters in M82. In fact, both the younger and older clusters of our sample
follow power-law forms, with the index being marginally different for the
two populations.
More importantly, the apparent turn-over masses for the two 
samples are similar, and 
are at least a factor of five lower than that reported by \citet{deG03a}. 
Bright clusters of M82 B make part of our sample, and hence these differences 
are intriguing.

The explanation for the different conclusions lies in the different adopted
ages: 100~Myr (ours) versus 1~Gyr \citep{deG03a}.
Available spectroscopic data favor ages between 50--350~Myr:
\citet{Smi06} derived an age of $\sim350$~Myr
for two knots (B2-1 and B1-1) in region M82 B, and 50--65~Myr for the knots 
F and L at similar galactocentric distance as region B, but to the southwest of 
the nucleus. Recent determination of ages for 7 more clusters 
by \citet{Kon07} also support these relatively younger ages.
Thus, an age of 1~Gyr seems extremely high for the cluster
population in M82, and the ages are likely to be closer
to the one we have adopted. The $B$-band luminosity drops by a factor of 8.5
due to passive evolution of SSPs between 100~Myr and 1~Gyr \citep{Gir02}.
Thus, the photometrically derived masses by \citet{deG03b} could have
been overestimated by a factor as large as this. If we take into account 
this factor, their turn-over mass values would be similar to
the ones we find in the present study. A lower turn-over mass dilutes the
case for a log-normal initial CMF.

\subsection{Disruption of clusters: mass and size dependence}

Whether low-mass clusters are preferentially disrupted with respect to
the high mass ones is the question waiting to be answered by
the ever growing data on extragalactic star clusters. So far, the best 
studied cases are M51 and the Large Magellanic Cloud (LMC). For M51 
clusters, \citet{Bas05} concluded that the disruption rate at early 
times ($\sim10$--30~Myr) is independent of cluster mass. In the LMC, 
\citet{deG07}  report evidence for mass-dependent disruption for masses 
below a few $\times10^3$~\msun. Our 
result showing a tendency for a flatter slope of the mass function
at older ages is an indication
that the disruption processes acting in the disk of M82 are mass-dependent.

For a given mass, clusters of larger radii are loosely bound, and 
it is easier to disrupt them. Hence, one would expect those clusters
to be missing in older cluster samples. \citet{Gie05} found size-dependent 
evolution for the M51 clusters. In the Antennae galaxy, \citet{Men05} 
have found a decrease of the average cluster size with age, as well. 
In the present study, we find that the average size
of the clusters is smaller for older clusters of mass $<10^5$~\msun.
Thus, evolutionary processes selectively disrupt loosely bound clusters.

\subsection{Physical processes driving the disruption}

In the introduction, we have mentioned several physical processes
that are capable of disrupting a star cluster. We refer 
the readers to the works of \citet{Spi87}, \citet{Fal01} and \citet{deG07}, 
for details on these physical processes.
At early times ($t\lesssim30$~Myr), disruption is caused mainly due to the
expulsion of the intra-cluster gas through supernova explosions.
This process is independent of cluster mass, as long as the stellar
IMF, or the mass fraction of the gas, are not systematically different for low
and high mass clusters. The other process that can change the slope
of the cluster mass function is the tidal shock experienced by
the clusters as they move in the gravitational field of a galaxy. 
According to \citet{Fal01}, this process becomes important 
after $\sim300$~Myr in normal galaxies. However, in the case of M82,
\citet{deG05} have estimated a disruption timescale as short as 30~Myr 
for a cluster of mass $10^4$~\msun\  at 1~kpc away from the center, with 
a dependence on mass that varies as $M^{0.6}$. Hence, the derived slope 
of $\alpha=1.8$ for the nuclear clusters, which are younger than 10~Myr, 
likely represents
the initial slope of the mass function. On the other hand, 
clusters in the disk 
of M82 are expected to suffer from the dynamical processes of cluster 
disruption. This kind of disruption process destroys first the larger 
clusters. The over-abundance of compact clusters in our sample of the older
clusters, as compared to the sample of younger clusters, is quite 
consistent with this idea. Thus our dataset support the relatively
short disruption timescale estimated for M82.

\subsection{Initial size distribution function}

Very little is known regarding the distribution of sizes of SSCs.
For clusters in M51, \citet{Bas05} found a power-law relationship 
with an index of $2.2\pm0.2$, which agrees well with that for the galactic 
globular clusters ($2.4\pm0.5$). Using more recent HST/ACS data
of M51, \cite{Sch07} also find a power-law form for the CSF, albeit 
over a much smaller range in sizes. 
In the case of M82, the power-law form is not 
evident in the observed distribution of sizes of bright clusters 
considered in the present study (Figure~\ref{fig_size_fun}): we should 
have detected large numbers of compact clusters, which is not the case. 
As compared to our sample in M82, the low-mass limit in the cluster
samples of M51 \citep{Bas05,Sch07} is an order of magnitude lower. 
If the low-mass clusters are systematically more compact than their
high-mass counterparts, then 
the observed difference in the functional forms of the
two galaxies can be understood in terms of the difference in the low-mass
limits. However, no clear mass-radius relation was found for M51 clusters.
Hence, the CSF for the bright cluster in M82 is different from 
that found in M51.
 
It is likely that this difference might be related to the relative youth
of the disk of M82. In a galaxy with clusters forming continuously over time,
the selective disruption of large clusters is expected to induce an
accumulation of clusters of the smaller sizes at any given cluster mass.
This cumulative process seems to have taken place in M51, while in M82 
we are likely witnessing the initial distribution of sizes.

\section{Summary}

In this study, we have carried out an objective search for star clusters 
on the HST/ACS images of M82 in filters F435W(B), F555W(V), and F814W(I). 
The search has led to the discovery of 393 clusters in the disk and 260 
clusters in the nuclear region. The magnitude and FWHM of these clusters 
were used to construct luminosity and size distribution functions.
We find that the luminosity function follows a power-law with an index of 2,
with an apparent turn-over at the faint end. Monte Carlo simulations 
carried out by us show that this turn-over is a 
consequence of incompleteness in the detection of faint clusters 
rather than a turn-over in the intrinsic luminosity of the population.
We used simple stellar population synthesis models to derive visual 
extinction values and photometric masses for the clusters, adopting
a uniform
age of 8~Myr for the nuclear clusters and 100~Myr for the disk ones. 
The resultant mass distribution functions for the nuclear and disk regions 
follow power-law functions, with a marginally steeper index value 
of 1.8 for the 
younger nuclear regions as compared to 1.5 for the older disk regions.
The cluster size distribution function, constructed for the clusters
brighter than the turn-over magnitude, follows a log-normal function
with its center at 10~pc FWHM ($R_{\rm eff}\sim4$~pc) for the most
massive clusters (mass $>10^5$~\msun).
For lower masses (mass=(0.2--1.0)$\times10^5$~\msun),
the center is marginally shifted to larger values for the
younger, and smaller values for the older clusters. 
This tendency implies that the extended low-mass clusters are
selectively destroyed during their dynamical evolution.
The marginally flatter 
slope of the mass function, and an over-abundance of
compact clusters in the older sample, are pointing towards a mass
and size dependent cluster disruption process at work in the disk of M82.

\acknowledgments

We would like to thank the Hubble Heritage Team at the Space Telescope
Science Institute for making the reduced fits files available to us.
We also thank an anonymous referee for many useful comments that have
lead to an improvement of the original manuscript. Additionally, we thank
Prof. Alessandro Bressan for useful comments on the manuscript.
This work is partly supported by CONACyT (Mexico) research grants
42609-F, 49942-F and 58956-F.

\clearpage

\begin{deluxetable}{lllllll}
\tablewidth{0pc}
\tablecaption{Exposure times and limiting magnitudes for stars and clusters
\tablenotemark{a}.}
\tablehead{
\colhead{Filter} & \colhead{Exp Time} & \colhead{m(star)}
& \multicolumn{2}{c}{m(cluster)} \\
\colhead{} & \colhead{seconds} &  & \colhead{compact} & \colhead{extended} 
}
\startdata
F435W (B) & 1600  & 25.67 & 21.76 & 23.97 \\
F555W (V) & 1360  & 25.18 & 21.27 & 23.48 \\
F814W (I) & 1360  & 24.24 & 20.32 & 22.53 \\
\hline
\enddata
\tablenotetext{a}{Limiting magnitudes are calculated using Gaussian intensity
profiles that enclose at least 10 pixels (stars) and 50 pixels (clusters) 
above the 5 $\sigma$ detection threshold. A FWHM of 2.1 pixels is assumed
for the stars, whereas cluster magnitudes are calculated using 3 pixels 
for compact, and 10 pixels for extended clusters. 
}
\end{deluxetable}

\begin{deluxetable}{lcccl} 
\tablewidth{0pc}
\tablecaption{Source detection statistics in M82}
\tablehead{
\colhead{Source type} & \colhead{$B$} & \colhead{$V$}
                   & \colhead{$I$} & \colhead{Selection Criteria} 
}
\startdata
All & 44274     &  82515   &  151565 & 5--10 $\sigma$/pixel \\
Extended & 83\%      &  77\%    &  60\%   & $3\le FWHM/{\rm pix}<30$\\
Candidates & 7632   &  7688   &  6307   & $3\le FWHM/{\rm pix}<30$ \& $area>50$ pix\\
Clusters\tablenotemark{a}     & 421 (263) &  456 (306) & 390 (208) & see $\S$2.1 \\
\hline
\enddata
\tablenotetext{a}{Numbers in the parenthesis are the disk clusters. }
\end{deluxetable}

\begin{deluxetable}{lccccccrcccl} 
\tabletypesize{\scriptsize}
\tablewidth{0pc}
\tablecaption{Photometric properties of detected star clusters in M82\tablenotemark{a}}
\tablehead{
\colhead{ID} & \colhead{$B$} & \colhead{$\sigma(B)$} & \colhead{$B-V$} 
             & \colhead{$V-I$} & \colhead{$A_{\rm v}$} & \colhead{$\sigma(A_{\rm v})$}  
   & \colhead{$M_B^0$} & \colhead{$\sigma(M_B^0)$} & \colhead{$\log(M\ast)$}  
             & \colhead{$\sigma(\log({\rm M\ast}))$}  & \colhead{Other ID}  \\
}
\startdata
   1N &  18.39 &   0.01 &  1.14 &  1.57 &   2.38 &   0.00 & $-12.63$ &   0.20 &   5.42 &   0.08  & H    \\ 
   2N &  18.43 &   0.01 &  0.69 &  0.93 &   1.01 &   0.22 & $-10.74$ &   0.36 &   4.67 &   0.14  & M35SE      \\
   3N &  18.68 &   0.01 &  1.27 &  2.09 &   3.19 &   0.62 & $-13.43$ &   0.86 &   5.75 &   0.35  & A1,M86SE   \\
   4N &  18.95 &   0.01 &  1.11 &  1.82 &   2.65 &   0.47 & $-12.43$ &   0.66 &   5.34 &   0.27  & \nodata    \\
   5N &  18.96 &   0.01 &  1.20 &  1.75 &   2.68 &   0.18 & $-12.47$ &   0.32 &   5.36 &   0.13  & M6NW       \\
   6N &  18.99 &   0.01 &  1.14 &  1.63 &   2.46 &   0.11 & $-12.14$ &   0.25 &   5.23 &   0.10  & M82SE      \\
   7N &  19.16 &   0.01 &  1.05 &  1.79 &   2.54 &   0.55 & $-12.08$ &   0.77 &   5.20 &   0.31  & M83SE      \\
   8N &  19.34 &   0.02 &  0.60 &  0.97 &   0.93 &   0.03 & $ -9.71$ &   0.20 &   4.26 &   0.08  & \nodata    \\
   9N &  19.35 &   0.01 &  0.73 &  1.42 &   1.65 &   0.53 & $-10.68$ &   0.74 &   4.64 &   0.30  & \nodata    \\
  10N &  19.45 &   0.00 &  1.30 &  2.16 &   3.32 &   0.68 & $-12.84$ &   0.94 &   5.51 &   0.38  & \nodata    \\
  11N &  19.47 &   0.01 &  0.92 &  1.38 &   1.86 &   0.09 & $-10.85$ &   0.24 &   4.71 &   0.09  & M20SE      \\
  12N &  19.50 &   0.01 &  1.35 &  2.02 &   3.21 &   0.35 & $-12.65$ &   0.51 &   5.43 &   0.20  & M77SE      \\
  13N &  19.57 &   0.01 &  0.73 &  1.07 &   1.24 &   0.06 & $ -9.90$ &   0.21 &   4.33 &   0.09  & M12SE      \\
  14N &  19.66 &   0.02 &  1.53 &  2.40 &   3.92 &   0.65 & $-13.45$ &   0.90 &   5.75 &   0.36  & M84SE      \\
  15N &  19.72 &   0.02 &  0.60 &  1.02 &   0.99 &   0.11 & $ -9.41$ &   0.25 &   4.14 &   0.10  & \nodata    \\
\hline
\enddata
\tablenotetext{a}{The Full table is available in the electronic version.}
\end{deluxetable}

\clearpage

\begin{figure}
\begin{center}
\figurenum{1}
\caption{Stamps of size $2.5^{\prime\prime}\times2.5^{\prime\prime}$ 
centered on the detected clusters in the disk of M82. The cluster $B$-band
brightness decreases with increasing identification number.
The image in which an object is detected with the highest signal-to-noise 
ratio is used to generate the stamps (i.e. the $B$ image in the first 
3 panels followed by $V$ and $I$ images). 
(The figure is presented as 5 image files named respectively as f1a.gif, 
f1b.gif, f1c.gif, f1d.gif and f1e.gif).
\label{fig_stamps}}
\end{center}
\end{figure}
\clearpage

\end{document}